\documentclass[fleqn,usenatbib]{mnras}

\usepackage{newtxtext,newtxmath}

\usepackage[T1]{fontenc}
\usepackage{float}

\DeclareRobustCommand{\VAN}[3]{#2}
\let\VANthebibliography\thebibliography
\def\thebibliography{\DeclareRobustCommand{\VAN}[3]{##3}\VANthebibliography}

\usepackage{graphicx}	
\usepackage{amsmath}	

\newcommand{\kms}{\,km\,s$^{-1}$}
\newcommand{\eso}{ESO\,138-G001}

\usepackage{xcolor}
\usepackage[normalem]{ulem}

\title[The NLR of ESO~138-G001]{The Narrow Line Region properties of ESO~138-G001 unveiled by SOAR/SIFS observations\thanks{Based on observations obtained at the Southern Astrophysical Research (SOAR) telescope, which is a joint project of the Ministério da Ciência, Tecnologia e Inovações do Brasil (MCTI/LNA), the US National Science Foundation’s NOIRLab, the University of North Carolina at Chapel Hill (UNC), and Michigan State University (MSU).}}

\author[A. Rodr\'{i}guez-Ardila et al.]{
A. Rodr\'{i}guez-Ardila,$^{1, 2}$\thanks{E-mail: aardila@lna.br}
D. May,$^{3,4}$
S. Panda,$^{1}$
M. A. Fonseca-Faria,$^{1}$ and L. Fraga$^{1}$
\\
$^{1}$Laborat\'orio Nacional de Astrofísica, Rua dos Estados Unidos, 154, Bairro nas Nações, Itajubá, MG, Brazil\\
$^{2}$Observat\'orio Nacional, Rua General Jos\'e Cristino 77, CEP 20921-400, S\~ao Crist\'ov\~ao, Rio de Janeiro, RJ, Brazil\\
$^{3}$Instituto de Astronomia, Geof\'isica e Ci\^encias Atmosf\'ericas, Universidade de S\~ao Paulo, 05508-090, S\~ao Paulo, SP, Brazil\\
$^{4}$ Gemini Observatory/NSF’s National Optical-Infrared Astronomy Research Laboratory, Casilla 603, La Serena, Chile
}

\date{Accepted XXX. Received YYY; in original form ZZZ}

\pubyear{2023}

\begin{document}
\label{firstpage}
\pagerange{\pageref{firstpage}--\pageref{lastpage}}
\maketitle

\begin{abstract}
We study in detail the inner $\sim$600~pc of the Seyfert~2 galaxy ESO\,138-G001 by means of the Soar Integral Field Spectrograph (SIFS) attached to the SOAR telescope. This source is known for displaying a very rich coronal line spectrum and a blob of high-excitation emission $\sim 3\arcsec$ SE of the active galactic nucleus (AGN). The nature of this emission has not been fully understood yet. The excellent spatial and spectral resolution of SIFS allows us to confirm that the bulk of the coronal line forest emission region is very compact,  of $\sim 0.8\arcsec$ in diameter, centred on the AGN and most likely powered by radiation from the AGN. 
In addition, evidence of a nuclear outflow, restricted to the inner 1$\arcsec$ centred at the nucleus is found based on the detection of broad components in the most important emission lines. The gas in the inner few tens of parsecs filters out the AGN continuum so that the NLR is basically illuminated by a modified SED. This scenario is confirmed by means of photoionisation models that reproduce the most important lines detected in the SIFS field-of-view. From the modelling, we also found that the black hole mass $M_{\rm BH}$ of the AGN is about 10$^{5.50}$~M$_{\odot}$, in agreement with previous X-ray observations.   The spectrum of the SE blob is dominated by emission lines of low- to mid-ionisation, with no hints of coronal lines. Our results show that it represents gas in the ionisation cone that is photoionised by the filtered central AGN continuum.
\end{abstract}




\section{Introduction}
The study of individual galaxies by means of Integral Field Unit (IFU) spectrographs has brought an impressive amount of knowledge about the central engine and the inner few hundred parsecs of Active Galactic Nuclei \citep[][to name a few]{Muller11, Mazzalay13, DMay18, Bianchin22, riffel22}. Much of the interest in these studies is because active galactic nuclei (AGN) commonly have outflows and jets, and these structures/phenomena strongly influence their surroundings. While direct jet–gas interactions affect the velocity field of the local medium, they also produce fast, auto-ionising shocks that can significantly influence (or even dominate) the observed emission-line strengths. In nearby AGNs, the warm ($T \sim~10^4$~K), ionised gas in the narrow-line region (NLR) can be resolved at scales of a few tens to a few hundred parsecs, allowing us to map the spatial variation of different gas properties and thus assessing if the observed emission-line spectrum is photoionised solely by the AGN or if the effect of wind/jets also contributes to the observed strength of the lines \citep{Emonts05, Ardila06, morganti13a, Fonseca-Faria+23}.

In this context,
\eso{} is a very interesting target to study in detail by means of IFU data. It is a low luminosity E/S0 galaxy hosting a Seyfert~2
nucleus  \citep{lipovetsky_1988, Schmitt_Storchi-Bergmann95}. The emission-line region of this AGN is resolved into several substructures distributed roughly along P.A. $\sim 135 \deg$,
the highest excitation gas  (highest [\ion{O}{iii}]/[\ion{N}{ii}]+H$\alpha$) laying 3" southeast of the nucleus.  
The bright, blue emission to the immediate southeast of the nucleus was studied by \citet{Ferruit00}, who argue that it is either scattered nuclear light from the Seyfert~2 nucleus or produced by young, hot stars. To the best of our knowledge, the nature of this spot is still open.

\eso{} is also known for displaying a wide range of ionisation in its optical spectrum, including lines of low-, intermediate- and high-ionisation stages. Indeed, due to the unusual strength and number of coronal lines detected, it was classified as a Coronal Line Forest AGN (CLiF AGN) by \citet{Rose15}.  
In these sources, the high-ionisation lines are enhanced relative to low- and mid-ionisation lines, allowing the detection of a forest of coronal lines rarely seen in non-CLiF AGN. \citet{Cerqueira-Campos_etal_2021} analysed optical and near-infrared (NIR) long-slit spectra of \eso{}. Their results show evidence of a hidden broad line region (BLR) along with an impressive number of coronal lines (CLs) of [Fe\,{\sc vii}], [Fe\,{\sc x}], and [Fe\,{\sc xi}] in the optical and up to [Fe\,{\sc xiii}] in the NIR.

In the X-ray region, \eso{} has been well studied and from its properties, it is considered a very peculiar source. For example, \citet{Collinge_Brandt00} using the Advanced Satellite for Cosmology and Astrophysics (ASCA) data suggests that it is a reflection-dominated source based on its hard spectrum and the prominent Fe~K$\alpha$ emission line. \citet{Piconcelli11} claimed that it is a Compton-thick galaxy. This result was later confirmed by \citet{DeCicco15}, who derived a column density of 10$^{25}$~cm$^{-2}$, excluding the presence of a direct view of the nuclear continuum even in the very hard X-ray band. According to these authors, the iron emission line at 6.4~keV is likely produced by moderately ionised iron (\ion{Fe}{xii} – \ion{Fe}{xiii}) while the soft X-ray emission is dominated by emission features identified in the high-spectral resolution RGS spectra. These emission features would be likely associated with the optical narrow line region (NLR), which in the case of \eso{} would be quite compact and unresolved even in the high-spatial-resolution $Chandra$ image.

To the best of our knowledge, no detailed studies covering the inner few arcsecs around the central engine by means of integral field unit (IFU) spectrographs have been done so far in ESO~138-G001 or other CLiF AGN. We notice that ESO~138-G001 was previously examined in the context of the survey dubbed S7 (The Siding Spring Southern Seyfert Spectroscopic Snapshot Survey) using the Wide Field Spectrograph, or WiFeS \citep{dopita+15}. However, the spatial resolution of this instrument is 1$\arcsec$, suitable to map the whole NLR but too low to examine the inner few hundred parsecs around the AGN.

The fact that ESO 138-G001 is the closest CLiF known to us makes it a perfect candidate to examine the true nature of this type of AGN. From the redshift of the emission lines, we derived a $z$ = 0.00914 (see Section~2), allowing us to resolve angular scales of $\sim$146~pc under seeing-limited conditions ($\sim$0.8$\arcsec$). Moreover, the nature of the faint extended blue emission to the immediate southeast of the nucleus is still a matter of debate. Is it a massive cloud ejected by the AGN? A prominent spot in the ionisation cone? Or is an interacting galaxy being cannibalised by \eso{}? Previous evidence from the bright X-ray emission of this source suggests that it is due to gas in the NLR. However, no detailed spectroscopy of the bright spot to the SE has yet been made. 

Here, we carry out a deep IFU study in \eso{} with the goal of analysing the gas emission, degree of ionisation, physical conditions of the NLR gas and the nature of the blue extended SE feature observed in this object. In Sect.~\ref{sect:obs}, we describe the observations, data reduction procedure and tools for removing the continuum emission and fitting the emission lines. In Sect.~\ref{sec:nlr} we study the spatial distribution of the ionised gas, the main spectroscopic features and the physical conditions of the NLR gas while in Sect.~\ref{sec:kin} the gas kinematics is presented. In Sect.~\ref{sec:cloudy-modelling} we modelled the observed spectra for the nuclear region and the SE blob feature using the spectral synthesis code, {\sc cloudy} \citep{Ferland_etal_2017}. Discussion and final remarks are in Sect.~\ref{sec:concl}.

Throughout the paper, $H_{\rm 0}$ = 70~km~s$^{-1}$~Mpc$^{-1}$, $\Omega_{\rm m}$ = 0.30, and $\Omega_{\rm vac}$ = 0.70, have been adopted. At the redshift of \eso{}, $z$ = 0.00914 \citep{Veron06}, 
the projected scale is 1$\arcsec$ =  182~pc  in the Galactic Standard of Rest reference system. 

\section{Observations, data reduction and data treatment}
\label{sect:obs}

\eso{} was observed on the night of August 8, 2017, using the 4.1-meter Southern Astrophysical Research (SOAR) Telescope equipped with the SOAR Integral Field Spectrograph \citep[SIFS,][]{lepine+03}. The Integral Field Unit (IFU) of SIFS is a lenslet-fibre system, which consists of a matrix containing 1300 lenslets and optical fibres arranged in a 50$\times$26 array. The optical fibres in the IFU output are vertically aligned as a pseudo slit at the bench spectrograph entrance. The instrument records 1300 spectra simultaneously on its e2v detector with 4096$\times$4112 square pixels of 15-micron size. The observations were conducted with the 700 l/mm grating using the 700B setup. The spatial sampling scale is 0.30~arcsec per fibre. This configuration resulted in data cubes with a field of view of (FoV) 15$\times$7.8~arcsec$^{2}$, with the spectra covering from 4200 to 7000~\AA. The average resolving power ($R$) achieved was 4200.

The observations of \eso{} were divided into four exposures of 1200 sec, resulting in a total integration time of 80 minutes. The standard star EG\,274, observed during the same night, was used as a flux calibrator. The data reduction was performed using a custom-developed pipeline implemented on {\sc Python}, including {\sc Astropy} \citep{astropy:2013, astropy:2018, astropy:2022}, {\sc ccdproc} \citep{ccdproc} and {\sc spectral$-$cube} libraries. The standard procedures for spectroscopic data reduction were followed, which include the subtraction of the bias level, trimming, flat-field correction, cosmic ray cleaning using {\sc PyCosmic} \citep{Husemann2012}, and wavelength calibration based on Ne-HgAr arc lamps.

SIFS was designed in a manner that the spectra of the fibres on the CCD are densely packed in the spatial direction, resulting in significant cross-talk between neighbouring fibres. To accurately extract the spectra from this instrument, an additional calibration method was implemented. SIFS contains a masking mechanism positioned in front of the IFU, which effectively isolates the spectra of specific fibres while blocking the light from neighbouring fibres. During the acquisition of a scientific target, the masking mechanism is held at a distance from the IFU, ensuring unobstructed access for data acquisition.

For this additional calibration, flat-field images are obtained with the mask in front of the IFU. Fibre identification and tracing are performed for each of these images. By moving the mask mechanism through 8 positions, all 1300 fibres can be sampled. A Gaussian fit for each fibre along the dispersion axis is executed, and the width and centre position of the Gaussians are registered. To perform the extraction of science spectra, a linear multiple Gaussian fit is employed in the spatial direction, with the width and centre positions of each fibre recorded from the previous step as fixed parameters. The only free parameters for this fit are the Gaussian amplitudes. The final step is the 2D image reconstruction and the construction of the SIFS data cube.

\subsection{Data treatment}
   \label{sec:dt}

A complete description of the data treatment applied to IFU data cubes may be found in the works of \citet{Menezes14, Menezes15, DMay16} and~\citet{ Menezes19}, each one followed by a science case. The following procedures were applied to all data sets: spatial re-sampling with quadratic interpolation to 0.15 arcsec (half of the detector pixel scale) and the Butterworth spatial filtering (with cut-off frequency \textsl{f} = 0.25 and 0.30 in the $x$ and $y$ directions, respectively), to remove high spatial frequencies without affecting the point spread function (PSF). The correction of the differential atmospheric refraction was not applied because the maximum centroid displacement along the spectra is 0.3 pixels (or 0.1 arcsec).
   
Finally, if a good PSF  model is known, the Richardson-Lucy PSF deconvolution method  \citep{Richardson72, Lucy74} may be applied to improve the spatial resolution of the data. The best scenario to apply a deconvolution method is met when a point-like source suitable to model the PSF is available in the same science observation. This is precisely our case because we detected a field star within the SIFS FoV at 6 arcsec SW from the nucleus.  The presence of such a star allows that each emission line analysed in this work has its own PSF, extracted in the same wavelength range. 
   
The PSF FWHM and Strehl ratio before and after the data treatment are shown in Table~\ref{table:psf}. An average improvement of 46\% is reached for the PSF FWHM and 246\% for the Strehl ratio. In fact, the typical Strehl ratio achieved is the same value expected for observations with the future updated Adaptive Optics facility SAMplus \citep{Moser18} for the SOAR Telescope. Another good example of an outstanding improvement is seen in the work of \citet{DMay18}, in the near-infrared. It is worth noticing that the fibre diameter in SIFS is 0.3$\arcsec$.  
Therefore, the best possible angular resolution is 0.6\,arcsec.  

    \begin{table}
    \begin{center}
    \caption[strehl]
    {The FWHM and Strehl ratios of the PSFs extracted along the data cube, before (B) and after (A) deconvolution.}
    \begin{tabular}{|c|c|c|c|c|}
    \hline \hline
    \multicolumn{5}{c}{PSF properties by emission line} \\ \hline
    & \multicolumn{2}{c}{FWHM (arcsec)} & \multicolumn{2}{c}{Strehl ratio (\%)} \\
    Line & B & A & B & A \\ \hline
    {[}\ion{O}{iii}{]} & 1.26 & 0.71 & 3.4 & 11.6 \\
    {[}\ion{Fe}{vii}{]} & 1.23 & 0.62 & 3.7 & 15.1 \\
    {[}\ion{Fe}{x}{]} & 1.21 & 0.65 & 3.8 & 12.9 \\
    H$\alpha$ \& {[}\ion{N}{ii}{]} & 1.21 & 0.63 & 3.9 & 12.3 \\
    {[}\ion{S}{ii}{]} & 1.14 & 0.63 & 3.9 & 12.8 \\
     
    \hline
    \end{tabular}
    \label{table:psf}
    \end{center}
    \end{table}

   \begin{figure*}
	 
	\resizebox{0.80\hsize}{!}{\includegraphics{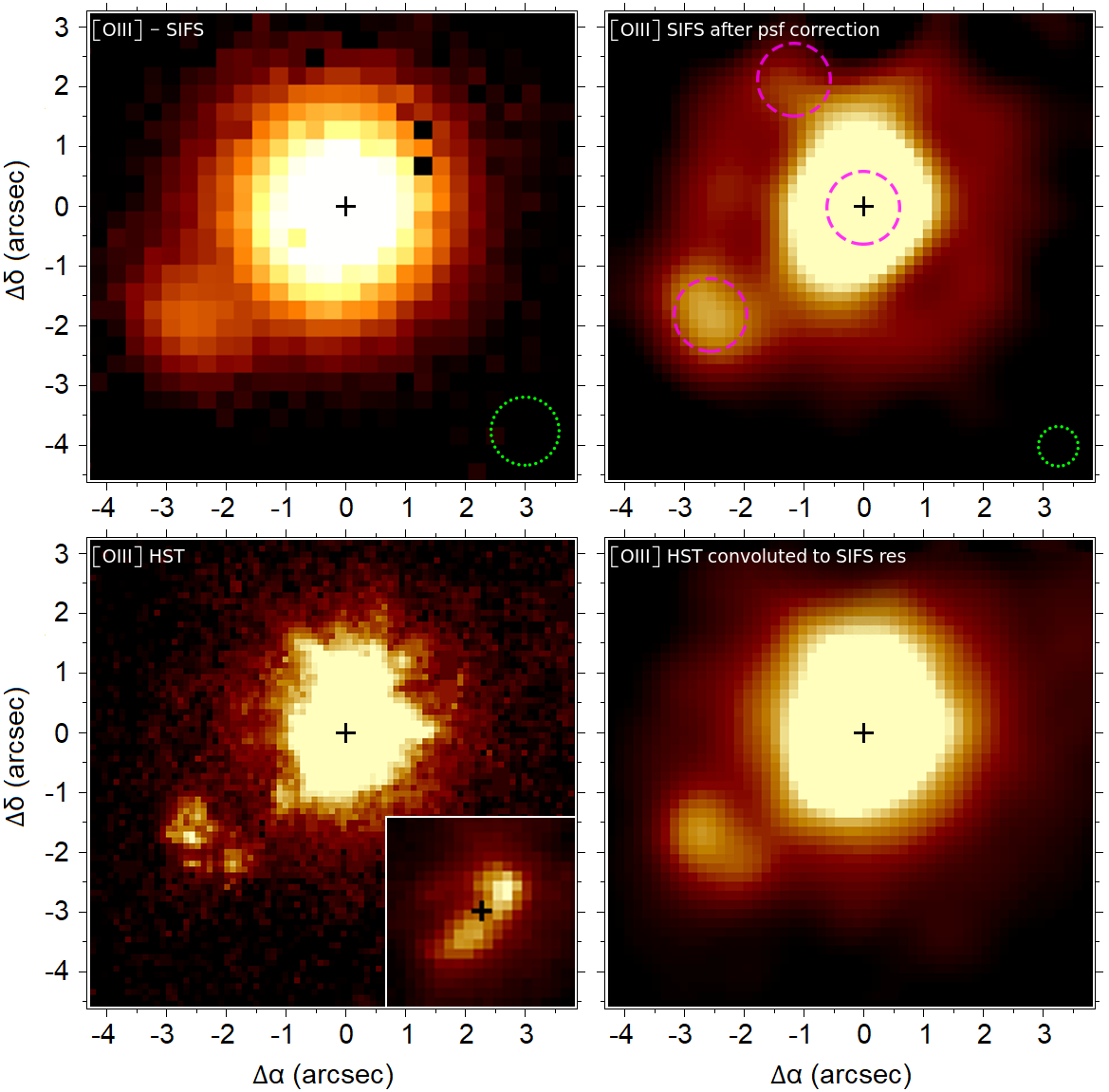}}
    \caption{Top:  [\ion{O}{iii}]~$\lambda$5007 image from SIFS, before the data treatment (left panel), and the treated image (right panel), with the green circles denoting the PSF FWHM for each of them ( 1.26 and 0.71 arcsec, respectively). The magenta circles - with an aperture radius of 0.6 arcsec - denote the extraction region of the spectra shown in Figure~\ref{fig:3spectra} with the emission line properties in Tables~\ref{tab:fluxnuc} and \ref{tab:fluxblobs}. Bottom: [\ion{O}{iii}]~$\lambda$5007 emission image from the HST with a small inset showing the details of the central part (left panel) and the same image convoluted with a 0.71 arcsec Gaussian PSF, binned to the same pixel size of SIFS-SOAR (right panel).  
    The two panels to the top and bottom right compare HST and SIFS with the same PSF. We notice that the faint extended emission is more prominent in the latter probably due to the SIFS's longer exposure time.   
    North is up and East is to the left. See the text for the discussion on the astrometry.}
    \label{fig:hst_sifs_before_after}
    \end{figure*}

To illustrate the effectiveness of the deconvolution method, we compare in Figure~\ref{fig:hst_sifs_before_after} the {[}\ion{O}{iii}{]} $\lambda$5007 \AA~ images from SIFS and HST. In order to perform a fair comparison we convolved the HST image with a PSF of  0.71 arcsec (see Table~\ref{table:psf}) and re-sampled the data to the same pixel size of SIFS (of 0.15$\arcsec$) after the data treatment. Looking at the SIFS images, the difference between the treated and non-treated data is noticeable (upper row)  
and has a quality similar to the convoluted HST image (bottom right panel). 

The treated SIFS image reveals features that were not seen before the data treatment, such as a clear morphology of the nuclear emission and two faint extended emission that seems to connect the NE and NW portion of the nucleus to the SE and SW regions, respectively, forming two "handle-like" features.  
The extended emission is not observed in the HST data, likely due to the shorter exposure time and number of individual exposures when compared to that of SIFS  (1$\times$800\,s for the former and 4$\times$1200\,s for the latter). Taking into account the telescope apertures, it translates to $\sim$18 times more collected light in the SIFS data compared to that of HST. In Sects.~\ref{sec:morf} and \ref{sec:kin} we confirm that the faint emission structure is real and kinematically distinct.
    
The astrometry between the HST and SIFS was done by defining the continuum centroid in the SIFS data - which coincides with that of the nuclear {[}\ion{O}{iii}{]}~$\lambda5007$ emission - and the centre of the convoluted {[}\ion{O}{iii}{]} image from the HST. This choice avoids the comparison of different nuclear positions because of the effect of distinct spatial resolutions, which is the case if one looks at the nuclear lobe-shape structure seen in the zoom of the raw HST image of Figure~\ref{fig:hst_sifs_before_after} (insert in the bottom left panel). If we had chosen to place the HST {[}\ion{O}{iii}{]} centre on its brightest nuclear spot, the comparison with the SIFS data would be no longer valid, as shown by the black cross in the inset.

\subsection{Continuum and emission line fitting}
    
In order to study the nature of the emission line gas present in the SIFS data cube it is necessary to remove the contamination caused by other components of the galaxy. In the wavelength region covered by SIFS, the observed continuum does not show any sign of absorption features or the presence of a featureless continuum, which led us to simply subtract the continuum by fitting a spline polynomial function of order 3.  

Moreover, we characterised the observed emission lines in terms of the integrated flux, full-width at half maximum (FWHM), and centroid position of the line peaks. To this purpose, we fitted Gaussian functions to individual lines or to sets of blended lines. This procedure was carried out using a set of custom 
scripts written in {\sc python} by our team.  
Usually, one or two Gaussian components were necessary to reproduce the observed emission line profiles. In this process, some 
constraints were applied. 
For instance, doublets such as [\ion{N}{ii}]~$\lambda\lambda6548,6583$ and [\ion{O}{iii}]~$\lambda\lambda$4959,5007  were constrained to their theoretical line flux ratio 1:3, to the same FWHM and to a fixed relative wavelength separation. Also, lines belonging to the same doublet but with no relative intensity constraints were set to have the same width and relative wavelength 
separation fixed by theory. This is the case of [\ion{S}{ii}]~$\lambda\lambda$6716,6731. Using that approach, the strongest emission lines detected in the IFU datacube  -- H$\beta$, [\ion{O}{iii}]$\,\lambda\lambda$4959,5007, [\ion{Fe}{vii}]$\,\lambda$6087, [\ion{O}{i}]$\,\lambda\lambda$6300,6364,  H$\alpha$, [\ion{N}{ii}]\,$\,\lambda\lambda$6548,6583, and [\ion{S}{ii}]\,$\lambda\lambda$6716,6731,  -- were fitted. The criterion for the best solution was the minimum value of the reduced ${\chi}^{2}$. 

\section{The narrow line region properties of ESO~138-G001}
\label{sec:nlr}
\subsection{Spatial distribution of the ionised gas}
\label{sec:morf}

\begin{figure*}
   \resizebox{0.99\hsize}{!}{\includegraphics{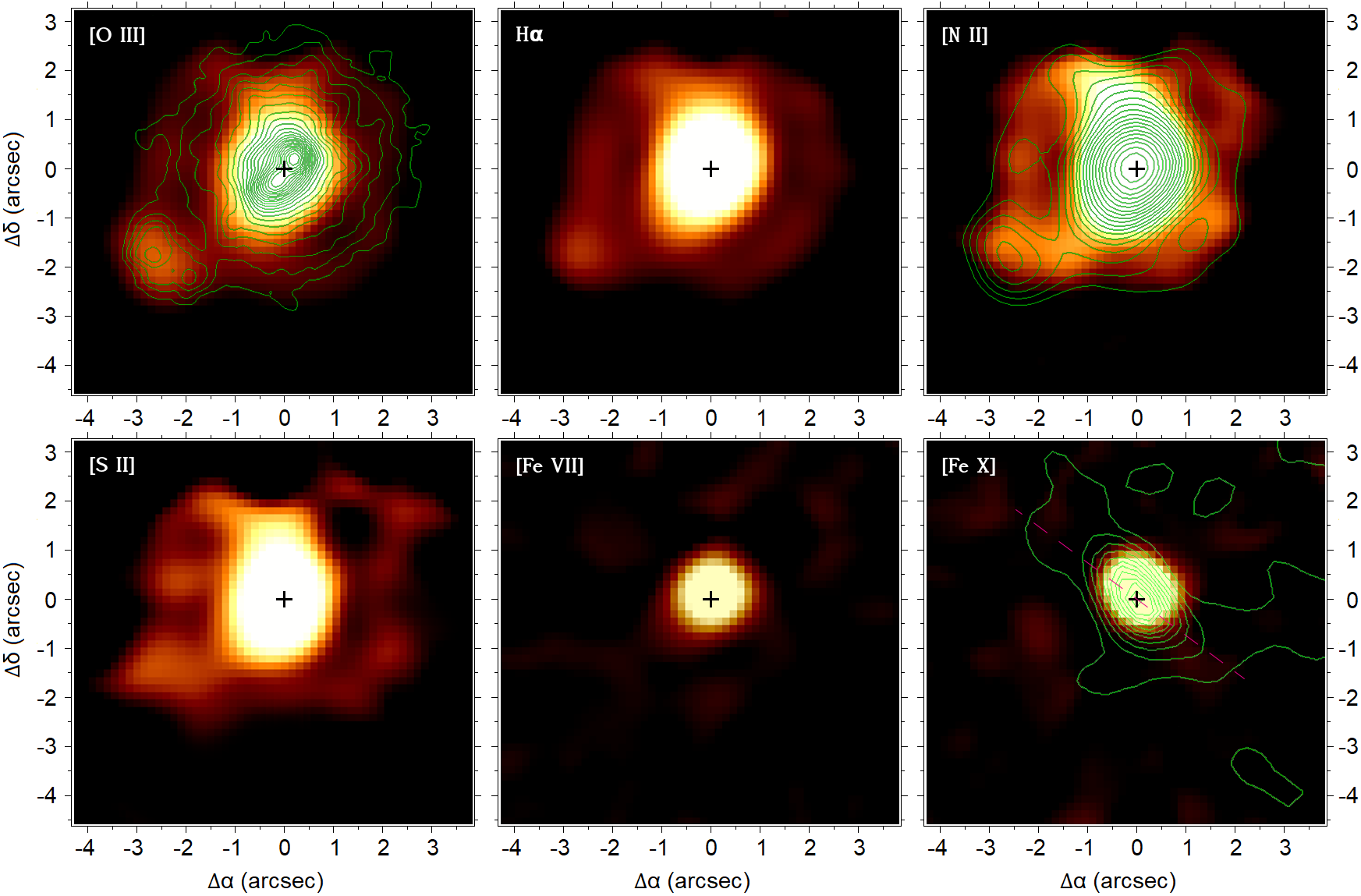}}
    \caption{Upper row: From left to right, emission line flux distribution for [\ion{O}{iii}]~$\lambda5007$, H$\alpha$~and [\ion{N}{ii}]~$\lambda6583$. Bottom row: the same for the [\ion{S}{ii}]~$\lambda\lambda$6716,6731, [\ion{Fe}{vii}]~$\lambda6087$ and [\ion{Fe}{x}]~$\lambda6374$ lines. Images are in a log scale. 
    Peaks in flux are 8.0, 5.0, 1.6, 0.6, 0.4 and 0.2 $\times 10^{-16}$~erg s$^{-1}$~cm$^{-2}$ \AA$^{-1}$, respectively. The green contours in the upper left are from the non-convoluted HST image and those in the upper right panel correspond to the SIFS [\ion{O}{iii}] emission.  The contours in the bottom right panel are from the radio jet, with the red dashed line showing the PA of that emission \citep{morganti99} In all panels, the North is up and the East is to the left.}
    \label{fig:lines}
\end{figure*}

With the purpose of studying the gas distribution of the narrow line region, we produce emission line flux maps for the most conspicuous lines detected in the SIFS cube. Examples of such maps are shown in Figure~\ref{fig:lines} for [\ion{O}{iii}]~$\lambda5007$, H$\alpha$~and [\ion{N}{ii}]~$\lambda6583$ (top row, from left to right panels, respectively) and [\ion{S}{ii}]~$\lambda\lambda6716,6731$, [\ion{Fe}{vii}]~$\lambda6087$, and [\ion{Fe}{x}]~$\lambda6374$ (bottom row, left to right panels, respectively). It can be seen that 
the gas emitting the former four lines is clearly extended. The bulk of the emission arises from the central 1.6$\arcsec$ region,  from where two faint extensions to the East and the West of the nucleus originate, forming two handles-like features. In contrast, the gas emitting the high-ionisation lines (i.e., [\ion{Fe}{vii}]~$\lambda6087$ and [\ion{Fe}{x}]~$\lambda6374$, hereafter [\ion{Fe}{vii}] and [\ion{Fe}{x}]) is fairly compact, with the maximum emission peaking very close to the AGN position. 

\begin{figure*}
   \resizebox{0.99\hsize}{!}{\includegraphics{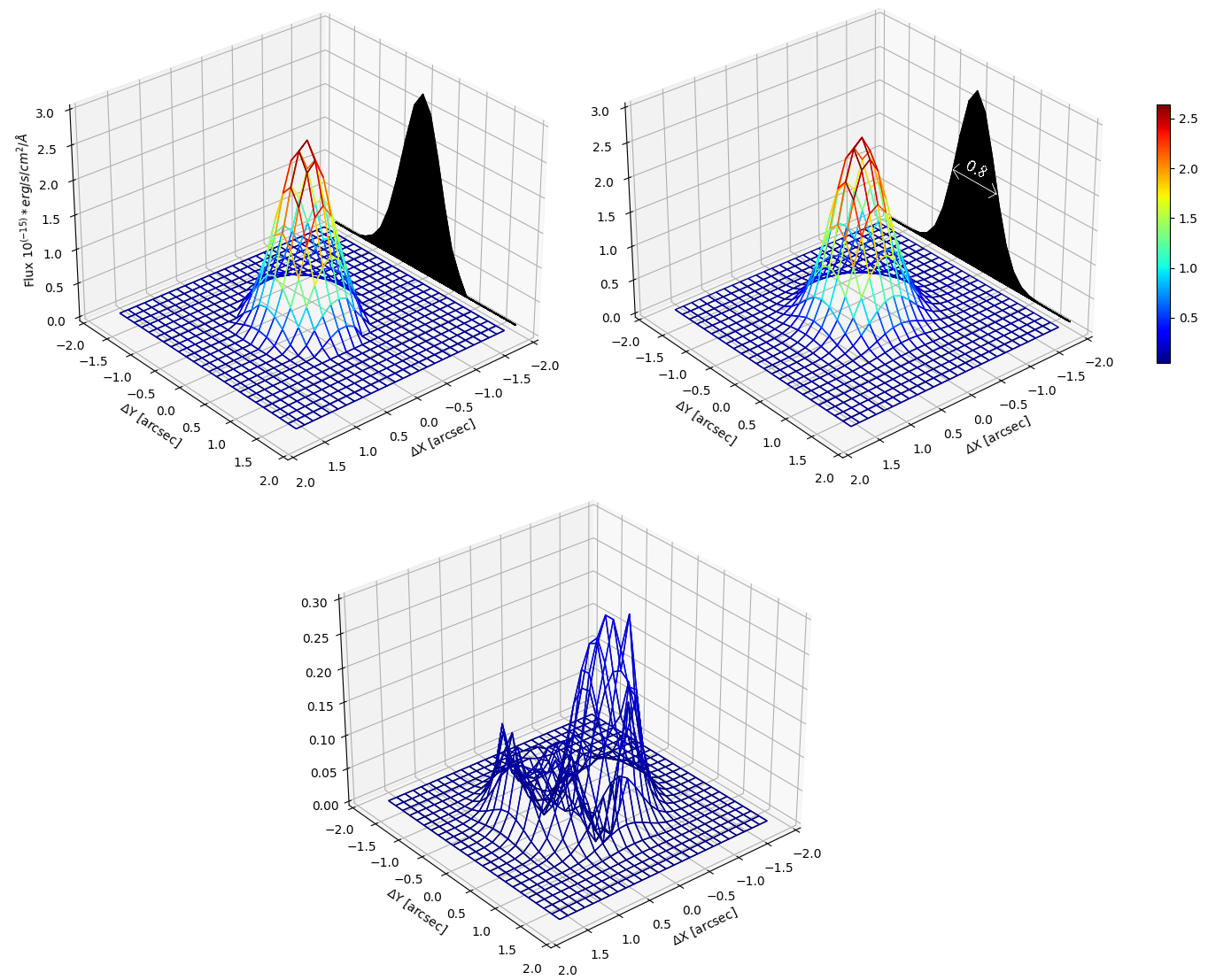}}
    \caption{ 
 Top left: Observed light profile of [\ion{Fe}{vii}] (left panel). Top right:  2D Gaussian model fit to the observations. Bottom panel: residual after subtraction of the model profile from the observations. For visualisation purposes, the scale of the Y-axis is an order of magnitude smaller than that of the two upper plots.  
    }
    \label{fig:fe7psf}
\end{figure*}

The upper left panel of Figure~\ref{fig:fe7psf} displays the observed light profile of [\ion{Fe}{vii}]. It can be seen that it is very peaked, with the bulk of the line emission restricted to the inner 0.8$\arcsec$. The right panel corresponds to the 2D Gaussian model fit to the observations. It has an FWHM = 0.8$\arcsec$ both in $x$ and $y$. The bottom panel displays the residual after subtraction of the model profile from the observations. The maximum residuals amount to 8\% of the peak signal. We consider that the Gaussian fit offers an excellent description of the [\ion{Fe}{vii}] light distribution, confirming that the emission is rather compact and constrained to a region of 146~pc (FWHM) in diameter. 

The [\ion{Fe}{x}] emission is also rather compact, similar to that of [\ion{Fe}{vii}]. The green contours overlaid to the [\ion{Fe}{x}] map represent the radio-emission as observed by \citet{morganti99}. The gas emission is slightly elongated along the direction of the radio jet. Moreover, the AGN position (identified with the "+" sign), does not coincide with the peak of the coronal emission but it is still within the uncertainties of the PSF. When applied a fit to the light distribution, we obtained a very similar result as that of [\ion{Fe}{vii}], except that the maximum residuals are somewhat higher (17\%) because of the slightly asymmetric profile. 

Two bright spots of  emission in  [\ion{O}{iii}], H$\alpha$, [\ion{N}{ii}], and [\ion{S}{ii}] are conspicuous in the maps displayed in Figure~\ref{fig:lines}. The first one is a prominent blob at $\sim3.2\arcsec$ SE of the AGN, first detected by \citet{Ferruit00} using HST/WFPC2 observations.  
The second one is a bright spot of emission $\sim2.6\arcsec$ to the NE of the AGN. It spatially coincides with the NE tip of the radio jet.  

From the emission line maps shown in Figure~\ref{fig:lines} (upper left panel), the match both in form and position of the SE blob between the SIFS and HST data in \eso{} is excellent. The blob displays a bean-like morphology, elongated along the NE-SW direction. From the HST contours (in green), two peaks of emission are observed, with the one at the NE being the brightest of the two.  Moreover, we notice a bridge connecting the central nuclear emission to the SW portion of the blob. The bridge is quite remarkable in the [\ion{N}{ii}] map and fainter in the [\ion{S}{ii}] emission.

\subsection{The emission line spectra in ESO~138-G001}
\label{sec:elf}

For the analysis that will be carried out in this subsection, we selected spectra of three representative regions of the IFU cube. They are marked in the upper right panel of Fig.~\ref{fig:hst_sifs_before_after} by the dashed magenta circles. The region at the centre of the image coincides with the AGN position. We also extracted spectra of the the NE-knot, a faint extension from the nucleus that shows up in low- to mid-ionisation gas; and the SE-blob - the most prominent circumnuclear structure whose nature has not been unveiled yet and we want to investigate here. The integrated region covered by each of the spectra comprises the size of the PSF before the deconvolution.   At the current angular resolution provided by SIFS, there are no other regions of particular interest regarding the emission lines in the IFU data cube.

\begin{figure*}
     \includegraphics[width=16cm]{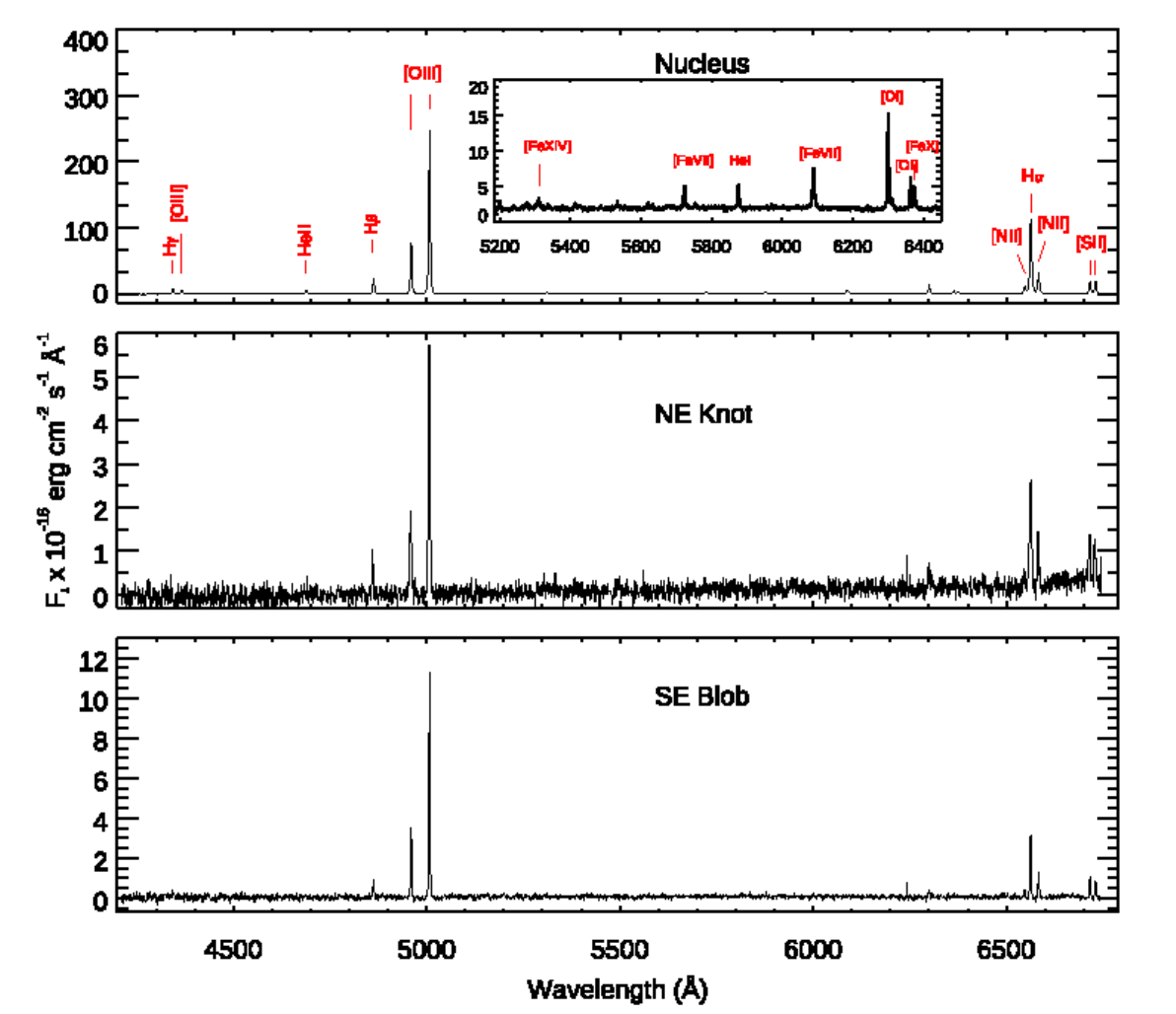}
    \caption{Integrated nebular emission line spectra of the nucleus (top panel), The NE-knot (middle panel), and the SE blob (bottom panel). In all cases, the observed spectra are corrected by the systemic velocity of the galaxy and have their continuum subtracted following the procedure described in the text.}
    \label{fig:3spectra}
\end{figure*}

The upper panel of Figure~\ref{fig:3spectra} corresponds to the AGN nuclear spectrum. The main emission lines detected (\ion{H}{i}, \ion{He}{i}, \ion{He}{ii}, [\ion{O}{iii}], [\ion{Fe}{vii}], [\ion{Fe}{x}], [\ion{Fe}{xiv}], [\ion{O}{i}],  [\ion{N}{ii}], and [\ion{S}{ii}]) are identified in red. The middle panel shows the spectrum of the NE-knot while the bottom panel displays the spectrum of the SE-blob. It can be clearly seen that in the two latter regions, only low- and mid-ionisation lines are present.
   
\begin{figure*}	 
	\includegraphics[width=12cm,angle=-90]{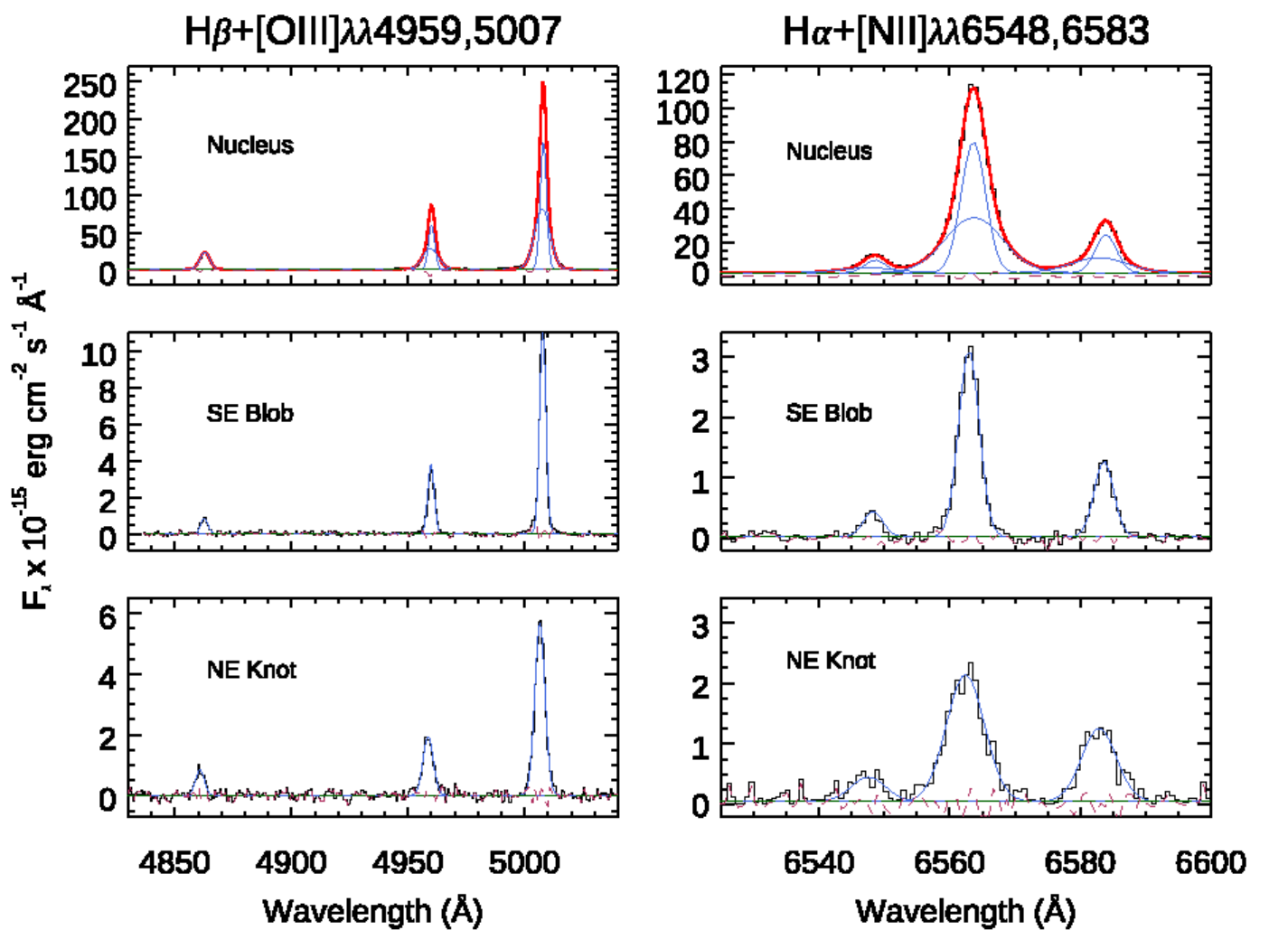}
	\caption{Examples of the Gaussian deblending applied to the H$\beta$+[\ion{O}{iii}] region (left panel) and H$\alpha$+[\ion{N}{ii}] (right panel). The observed spectrum is in black, individual Gaussian components are in blue while the total fit is in red. The green line is the continuum fit. The brown dashed line is the residual left after subtraction from the data set in the best-fit model.}
    \label{fig:gauss_fit}
\end{figure*}

Figure~\ref{fig:gauss_fit} shows examples of the line fitting procedure applied to the different lines of the spectra. The left column displays the results for H$\beta$+[\ion{O}{iii}] while the right column those for H$\alpha+$[\ion{N}{ii}]. The top, middle and bottom rows correspond to the fits carried out to the lines at the AGN position, the SE blob and the NE knot, respectively. We notice that only at the AGN position, two Gaussian components were necessary to model these lines. At other locations, the spectral lines were well reproduced with a single Gaussian component. 

Table~\ref{tab:fluxnuc} lists the emission lines detected at the nuclear position (col.~1), the centroid position of the line (col.~2), the measured integrated fluxes (col.~3), and the FWHM (col.~4). Table~\ref{tab:fluxblobs} is similar to Table~\ref{tab:fluxnuc} but now for the SE blob and NE knot of extended emission already described.

\begin{table}
\caption{Emission line fluxes (in units of 10$^{-16}$~erg~cm$^{-2}$\,s$^{-1}$) and full-width at half maximum (FWHM) of the lines (in km\,s$^{-1}$) measured at the AGN position in \eso{}.}
\label{tab:fluxnuc}
\begin{tabular}{l@{\hspace*{30pt}}l@{\hspace*{30pt}}l@{\hspace*{30pt}}l}
\hline
\hline
Line  &	   $\lambda$ (\AA) &		Flux &	   		FWHM \\
\hline
H$\gamma$ &	4343.31	&	41.3$\pm$1.05	&	356	\\
{[}\ion{O}{iii}{]}	&	4366.35	&	35.0$\pm$1.17	&	397	\\
\ion{He}{i}	&	4474.74	&	5.5$\pm$1.76	&	593	\\
\ion{He}{ii}	&	4687.78	&	33.6$\pm$0.94	&	348	\\
{[}\ion{Fe}{iii}{]}	&	4659.98	&	3.98$\pm$0.88	&	328	\\
{[}\ion{Ar}{iv}{]}	&	4713.13	&	3.9$\pm$0.80	&	291	\\
{[}\ion{Ar}{iv}{]}	&	4741.93	&	5.44$\pm$0.80	&	291	\\
H$\beta$	&	4862.8	&	116.97$\pm$0.80	&	281	\\
{[}\ion{O}{iii}{]}	&	4960.09	&	195.6$\pm$0.52	&	164	\\
{[}\ion{O}{iii}{]}$^*$	&	4959.59	&	224.76$\pm$1.28	&	458	\\
{[}\ion{O}{iii}{]}	&	5008.02	&	573.16$\pm$0.52	&	164	\\
{[}\ion{O}{iii}{]}$^*$	&	5007.52	&	658.55$\pm$1.28	&	458	\\
{[}\ion{Fe}{vi}{]}	&	5148.69	&	5.98$\pm$1.50	&	475	\\
{[}\ion{Fe}{vii}{]}	&	5160.67	&	8.595$\pm$1.07	&	332	\\
{[}\ion{Fe}{vi}{]}	&	5178.35	&	3.86$\pm$0.68	&	198	\\
{[}\ion{N}{i}{]}	&	5199.91	&	6.77$\pm$0.97	&	296	\\
{[}\ion{Fe}{vii}{]}	&	5278.74	&	4.49$\pm$0.86	&	283	\\
{[}\ion{Fe}{xiv}{]}	&	5303.61	&	4.00$\pm$0.86	&	282	\\
{[}\ion{Ca}{v}{]}	&	5311.07	&	8.45$\pm$0.86	&	281	\\
{[}\ion{Fe}{vi}{]}	&	5336.48	&	4.23$\pm$1.10	&	363	\\
\ion{He}{ii}	&	5413.3	&	3.91$\pm$0.87	&	280	\\
{[}\ion{Fe}{vi}{]}	&	5426	&	3.31$\pm$1.45	&	478	\\
{[}\ion{Fe}{vi}{]}	&	5533.81	&	7.93$\pm$1.75	&	567	\\
{[}\ion{Fe}{vi}{]}	&	5630.81	&	3.28$\pm$1.36	&	429	\\
{[}\ion{Fe}{vi}{]}	&	5677.23	&	5.18$\pm$1.86	&	590	\\
{[}\ion{Fe}{vii}{]}	&	5722.92	&	21.35$\pm$1.10	&	339	\\
{[}\ion{N}{ii}{]}	&	5754.88	&	3.19$\pm$0.81	&	241	\\
\ion{He}{i}	&	5876.25	&	16.82$\pm$0.83	&	244	\\
{[}\ion{Fe}{vii}{]}	&	6088.66	&	19.31$\pm$0.79	&	222	\\
{[}\ion{Fe}{vii}{]}$^*$	&	6089.43	&	23.51$\pm$1.99	&	589	\\
{[}\ion{O}{i}{]}	&	6300.61	&	40.78$\pm$0.60	&	156	\\
{[}\ion{O}{i}{]}$^*$	&	6300.61	&	28.03$\pm$1.49	&	421	\\
{[}\ion{S}{iii}{]}	&	6312.62	&	7.27$\pm$0.92	&	254	\\
{[}\ion{O}{i}{]}	&	6364.1	&	12.8$\pm$0.60	&	156	\\
{[}\ion{O}{i}{]}$^*$	&	6364.1	&	8.8$\pm$1.49	&	421	\\
{[}\ion{Fe}{x}{]}	&	6375.04	&	17.04$\pm$0.93	&	253	\\
{[}\ion{N}{ii}{]}	&	6548.59	&	30.67$\pm$0.57	&	158	\\
{[}\ion{N}{ii}{]}$^*$	&	6547.52	&	31.83$\pm$1.51	&	454	\\
H$\alpha$	&	6563.63	&	366.35$\pm$0.66	&	187	\\
H$\alpha^*$	&	6563.72	&	363.03$\pm$1.56	&	469	\\
{[}\ion{N}{ii}{]}	&	6583.92	&	92.02$\pm$0.57	&	158	\\
{[}\ion{N}{ii}{]}$^*$	&	6582.85	&	95.52$\pm$1.51	&	454	\\
\ion{He}{i}	&	6678.7	&	4.76$\pm$1.12	&	228	\\
{[}\ion{S}{ii}{]}	&	6716.9	&	32.86$\pm$0.67	&	122	\\
{[}\ion{S}{ii}{]}$^*$	&	6716.39	&	67.7$\pm$1.49	&	309	\\
{[}\ion{S}{ii}{]}	&	6731.43	&	35.3$\pm$0.67	&	122	\\
{[}\ion{S}{ii}{]}$^*$	&	6730.39	&	72.28$\pm$1.49	&	309	\\
\hline
\multicolumn{4}{l}{$^*$ Broad component of the line}
\end{tabular}
\end{table}

\begin{table}
\caption{Emission line fluxes (in units of 10$^{-16}$~erg~cm$^{-2}$\,s$^{-1}$) and full-width at half maximum (FWHM) of the lines (in km\,s$^{-1}$) measured at the SE Blob and NE knot in \eso{}.}
\label{tab:fluxblobs}
\begin{tabular}{l@{\hspace*{30pt}}l@{\hspace*{30pt}}l@{\hspace*{30pt}}l}
\hline
\hline
Line  &	   $\lambda$ (\AA) &		Flux &	   		FWHM \\
\hline
\multicolumn{4}{c}{SE Blob} \\
\hline
H$\beta$	&	4862.65	&	3.16$\pm$0.41	&	189	\\
{[}\ion{O}{iii}{]}	&	4959.83	&	13.05$\pm$0.38	&	169	\\
{[}\ion{O}{iii}{]}	&	5007.76	&	38.23$\pm$0.38	&	169	\\
{[}\ion{O}{i}{]}	&	6300.3	&	1.95$\pm$0.52	&	253	\\
{[}\ion{N}{ii}{]}	&	6548.29	&	1.55$\pm$0.32	&	143	\\
H$\alpha$	&	6562.95	&	12.21$\pm$0.33	&	153	\\
{[}\ion{N}{ii}{]}	&	6583.62	&	4.66$\pm$0.32	&	143	\\
{[}\ion{S}{ii}{]}	&	6716.82	&	3.89$\pm$0.31	&	136	\\
{[}\ion{S}{ii}{]}	&	6731.18	&	3.1$\pm$0.31	&	136	\\
\hline
\multicolumn{4}{c}{NE knot} \\
\hline
H$\beta$	&	4860.79	&	3.86$\pm$0.79	&	251	\\
{[}\ion{O}{iii}{]}	&	4958.6	&	10.59$\pm$0.93	&	295	\\
{[}\ion{O}{iii}{]}	&	5006.54	&	31.04$\pm$0.93	&	295	\\
{[}\ion{O}{i}{]}	&	6299.79	&	3.13$\pm$0.92	&	232	\\
{[}\ion{N}{ii}{]}	&	6547.11	&	2.84$\pm$1.16	&	252	\\
H$\alpha$	&	6562.33	&	17.06$\pm$1.37	&	301	\\
{[}\ion{N}{ii}{]}	&	6582.44	&	8.53$\pm$1.16	&	252	\\
{[}\ion{S}{ii}{]}	&	6714.91	&	5.53$\pm$1.07	&	224	\\
{[}\ion{S}{ii}{]}	&	6729.27	&	4.32$\pm$1.07	&	224	\\
\hline
\end{tabular}
\end{table}

The presence of strong high-ionisation lines in ESO\,138-G001 not only in the optical region but also in the near-infrared is outstanding and was previously reported by other authors   \citep{Alloin92, Cerqueira-Campos_etal_2021}.  
Our SIFS observations allow us to set a limit to the size of the coronal emission region in this AGN for the first time.
We found that the bulk of the coronal emission is produced in a region size of $\sim0.8\arcsec$ (FWHM) around the AGN.
This is consistent with the spectra shown in the middle and bottom panels of Figure~\ref{fig:3spectra}. They correspond to the two off-nuclear blobs $> 2\arcsec$ from the AGN. In both cases, the spectra are dominated by low- to mid-ionisation lines, with no hint of the presence of coronal lines. 

The bottom panel of Figure~\ref{fig:3spectra} displays the spectrum corresponding to the blob at $\sim2.6\arcsec$ SE of the AGN. According to \citet{Ferruit00},   
it displays the highest [\ion{O}{iii}]/([\ion{N}{ii}] $+$ H$\alpha$) ratio in the galaxy. From the values of fluxes listed in Tables~\ref{tab:fluxnuc} and~\ref{tab:fluxblobs} we measured a ratio of 2.25 at the blob while at the nucleus and the NE-knot we derived values of 1.34 and 1.21, respectively, in very good agreement to previous findings.  \citet{Ferruit00} argued that the blob is either scattered
nuclear light from the Seyfert~2 nucleus or produced by young, hot stars.  The lack of absorption features or a blue continuum emission associated with a young stellar population allows us to discard this latter possibility.

\subsection{Physical conditions of the NLR}
\label{phys_cond}

\subsubsection{Internal extinction distribution}
\label{ext}

The host galaxy of \eso{} is morphologically classified as E/S0  \citep{lauberts1982, Alloin92}. Therefore,  it is expected that dust
obscuration does not play an important role in the interpretation of
the galaxy structure. 
In order to confirm this hypothesis,
it is necessary to quantify the amount
of extinction towards the nucleus and in the circumnuclear region of 
the AGN. 

The presence of the H$\alpha$ and H$\beta$ lines in the spectra of the three main regions studied here  allowed us
to evaluate, for the first time, the intrinsic extinction affecting the nuclear and off-nuclear gas within the central $\sim$1~kpc by means of the Balmer decrement.

To this purpose, we used Equation~\ref{eq1}, which makes use of the observed H$\alpha$/H$\beta$ emission line flux ratio and the Cardelli, Clayton, \& Mathis -- CCM extinction law \citep{ccm89}. An intrinsic value of 3.1 for the above ratio was adopted,   assuming case B recombination. It is more suitable to the typical gas
densities found in AGN \citep{osterbrock06} and accounts
for the effect of collisional excitation. Equation~\ref{eq1} was applied to the three emission regions that are the focus of this analysis. We assumed that AGN photoionisation dominates in the regions of interest to this work.
 
\begin{equation}
    E(B-V)_{\rm H\alpha / H_\beta} = -2.31 \times log \left(\frac{3.1}{H\alpha / H\beta}\right)
	\label{eq1}
\end{equation} 
         
In Eq.~\ref{eq1}, H$\alpha$ and  H$\beta$ are the  observed
 emission lines fluxes of these two lines,
listed in Tables~\ref{tab:fluxnuc} and~\ref{tab:fluxblobs}. It is important to notice that only the flux of the narrow component of the two lines was employed. This is because the broad component was not identified in H$\beta$. Moreover, because that component is also observed in [\ion{O}{iii}] and other forbidden lines, it is likely associated with a nuclear outflow component. We attribute the lack of a broad emission in H$\beta$  to the lower S/N of that line when compared to H$\alpha$.  
The second column of Table~\ref{tab:properties}  lists the values of E(B-V) found for the nuclear and off-nuclear apertures.

It can be seen that \eso{} is, overall, characterised by low values of extinction. The nucleus is essentially dust-free while the SE blob displays an E(B-V) of only 0.22$\pm$0.1 mag. The amount of dust increases slightly to the NE when it reaches 0.35$\pm$0.13 mag. These results are in accord with the morphological type of the host galaxy, as mentioned at the beginning of this section. \citet{Alloin92}  
reported a Balmer decrement H$\alpha$/H$\beta$ of 4.25, which translates into an E(B-V) = 0.32 using the same extinction law and intrinsic Balmer decrement as here. Recently, \citet{Cerqueira-Campos_etal_2021} report an E(B-V) of 0.46$\pm$0.01. These latter values are slightly larger than the ones found here, but considering our ability to spatially resolve the different emission regions in the central few arcsecs of the AGN, not possible in the previous works, they overall agree.

\subsubsection{The electron density} \label{sec:density}

We determined the gas density by means of the [\ion{S}{ii}]~$\lambda$6716/$\lambda$6731 emission line flux ratio. To this purpose, we employed Equation~3 of \citet{Proxauf_Kimeswenger14},  which is an upgrade from the equations found by \citet{osterbrock06}. 
In this process, we assume an electron temperature $T_{\rm e}$ = 10000~K. For the nuclear aperture, we employed the flux values of the narrower components of both lines. Column~3 of Table~\ref{tab:properties} lists the values found. Our results show that the gas density at the nucleus reaches the highest value, amounting to 570$\pm$120~cm$^{-3}$. It is very typical of the density found in the nuclear region of most AGN \citep{DMay18,kakkad18, Cerqueira-Campos_etal_2021}. We notice that if we employ the flux values found for the broader components of the [\ion{S}{ii}] lines, we obtain essentially the same electron density, as the line ratio between $\lambda$6716/$\lambda$6731 is the same for both narrow and broad components.
 
\begin{table}
\caption{Physical properties of the emission gas in \eso{}.}
\label{tab:properties}
\begin{tabular}{lcccc}
\hline
\hline
Region  &	   E(B-V) &		$n_{\rm e}$ [\ion{S}{ii}] & $n_{\rm e}$ [\ion{Ar}{iv}] & $T_{\rm e}$\\    
& (mag) & (cm$^{-3}$) & (cm$^{-3}$) & ($^{\rm o}$K)   \\
\hline
Nucleus & 0.0 & 570$\pm$120 & 4784$\pm$1170 & 21044$\pm$2300 \\
SE Blob & 0.22$\pm$0.1 & 123$\pm$90 & ... & ... \\
NE knot & 0.35$\pm$0.13 & 105$\pm$97 & ... & ... \\
\hline
\end{tabular}
\end{table}

In addition to the sulphur lines, the gas density may also be determined by means of the  [Ar\,{\sc iv}]~$\lambda\lambda$4711,4740 doublet. These two lines are indeed detected in the nuclear spectrum of \eso{}. Because of the higher ionisation potential of the \ion{Ar}{iv}  ion (40.7~eV),  the gas density determined by these lines is more representative of the density in the region that also produces the [\ion{O}{iii}] lines compared to $n_{\rm e}$([\ion{S}{ii}]) lines \citep{Proxauf_Kimeswenger14}. Here, we are assuming a stratification of the gas ionisation and density, in the sense that higher-ionisation lines are produced in regions of higher gas density and vice-versa. With this in mind, we use Equation~5 of \citet{Proxauf_Kimeswenger14}, which is valid for a gas temperature in the interval  8000 $\leq T_{\rm e}\;({\rm K}) \leq$ 26000, and the fluxes of the $\lambda\lambda$4711,4740 lines listed in Table~\ref{tab:fluxnuc}. Our result is shown in the fourth column of Table~\ref{tab:properties}. It points out to a $n_{\rm e}$ value of 4784$\pm$1170~cm$^{-3}$, that is, nearly an order of magnitude larger than the density found by means of [\ion{S}{ii}].

The gas temperature $T_{\rm e}$ was determined by means of the [\ion{O}{iii}]~(4959~\AA\ + 5007~\AA / 4363~\AA)\ ratio and Eq.~1 of \citet{Proxauf_Kimeswenger14}. A value of 21044$\pm$2300 is obtained. It is within the interval of temperature assumed for the density determination using the [Ar\,{\sc iv}]. We report $T_{\rm e}$ only for the galaxy nucleus because $\lambda$4363 was only detected at that position.

The results shown in this section indicate that a large interval in gas density is present in the nuclear region of \eso{}, ranging from $\sim$600 to $\sim$4800 ~cm$^{-3}$. Similar findings were related by \citet{DMay18} and \citet{Bianchin22} in galaxies hosting an AGN. The temperature found for the gas supports photoionisation by the central source.

\section{Kinematics of the NLR emission gas}
\label{sec:kin}

\begin{figure*}
   \resizebox{0.90\hsize}{!}{\includegraphics{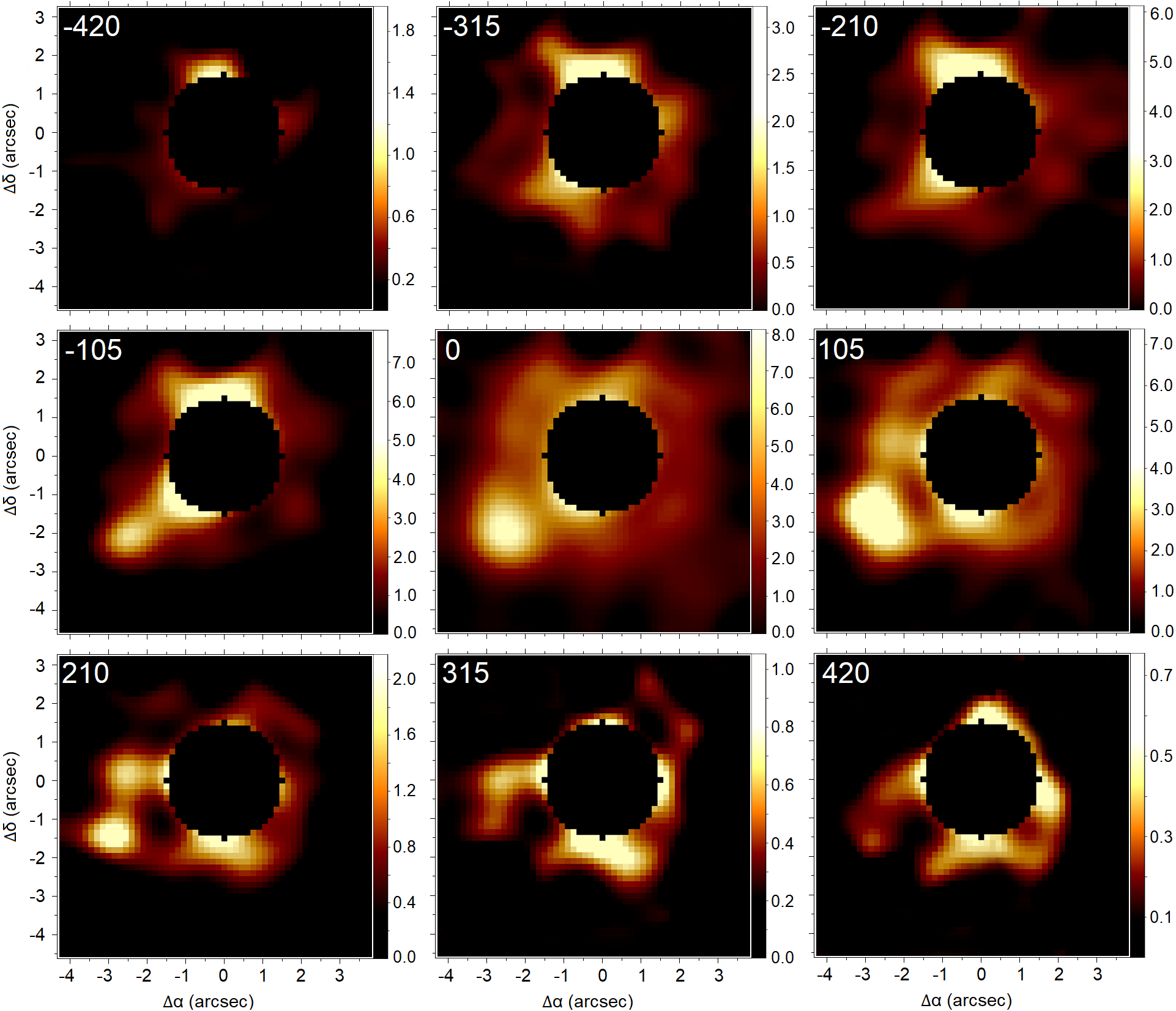}}
    \caption{Velocity channel maps for the [\ion{O}{iii}]~$\lambda5007$ emission line in  ESO~138-G001 derived from the IFU datacubes analysed in this work. The flux is in units of $10^{-16}$~erg s$^{-1}$~cm$^{-2}$ and, in each panel, the number at the left upper corner represents the velocity bin of the corresponding slice. We masked the nucleus (black region at the AGN position) in order to see faint extended features as they are $\sim$2 dex dimmer than the  AGN. In all panels, North is up and East is to the left.}
    \label{fig:oiii_channel_maps}
\end{figure*}

\begin{figure*}
   \resizebox{0.90\hsize}{!}{\includegraphics{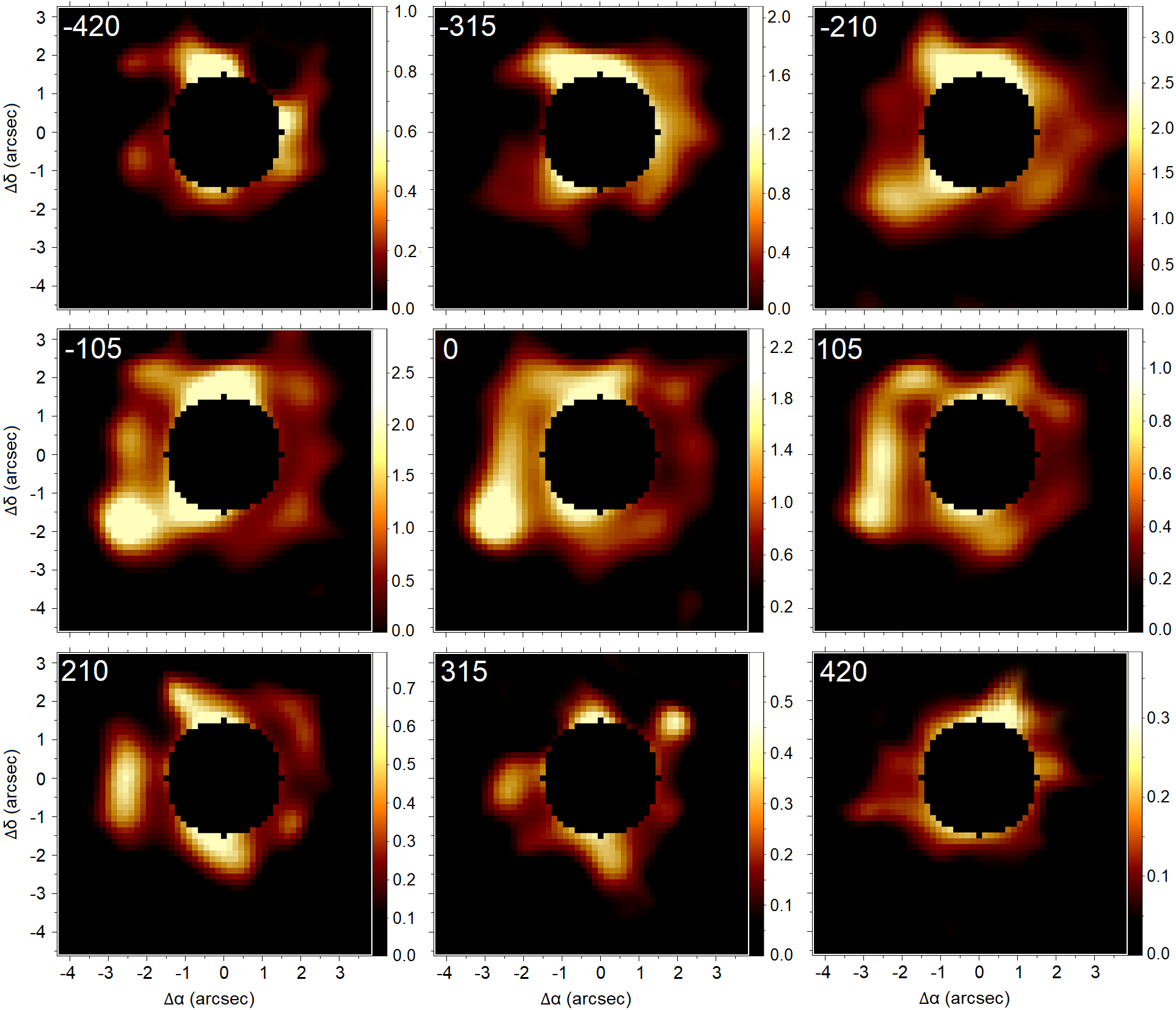}}
    \caption{Same as Figure~\ref{fig:oiii_channel_maps} for the H$\alpha$ emission line. The flux is in units of $10^{-16}$~erg s$^{-1}$~cm$^{-2}$.}
    \label{fig:ha_channel_maps}
\end{figure*}

Key information on gas kinematics can be obtained by slicing the observed line profiles in velocity bins called channel maps. The location where
blueshifted or redshifted gas is produced can be investigated
through this technique. With this in mind, channel maps for the
[\ion{O}{iii}]~$\lambda$5007 and the H$\alpha$ lines were constructed after subtracting the systemic velocity of the galaxy. The results are shown
in Figs~\ref{fig:oiii_channel_maps} and~\ref{fig:ha_channel_maps}, respectively. The central black circle masks the emission from the AGN so that fainter features are enhanced.

In both maps, it can be noticed that the most conspicuous extended features are the SE blob and the NE knot of emission. The former is clearly seen in the channels -105~km\,s$^{-1}$ to 210~km\,s$^{-1}$  for [\ion{O}{iii}] and -105~km\,s$^{-1}$ to 105~km\,s$^{-1}$ for H$\alpha$. The latter is most prominent in the channels -315 to 0~\kms\ for both lines. Regarding the SE blob, the brightest emission is centred at the systemic velocity of the galaxy, suggesting that the bulk of this gas is at rest with respect to the galaxy nucleus. If this blob represents gas ejected by the nucleus, Figures~\ref{fig:oiii_channel_maps} and~\ref{fig:ha_channel_maps} indicate that it is moving nearly perpendicular to the line of sight, in the plane of the sky.

\begin{figure*}
   \resizebox{0.90\hsize}{!}{\includegraphics{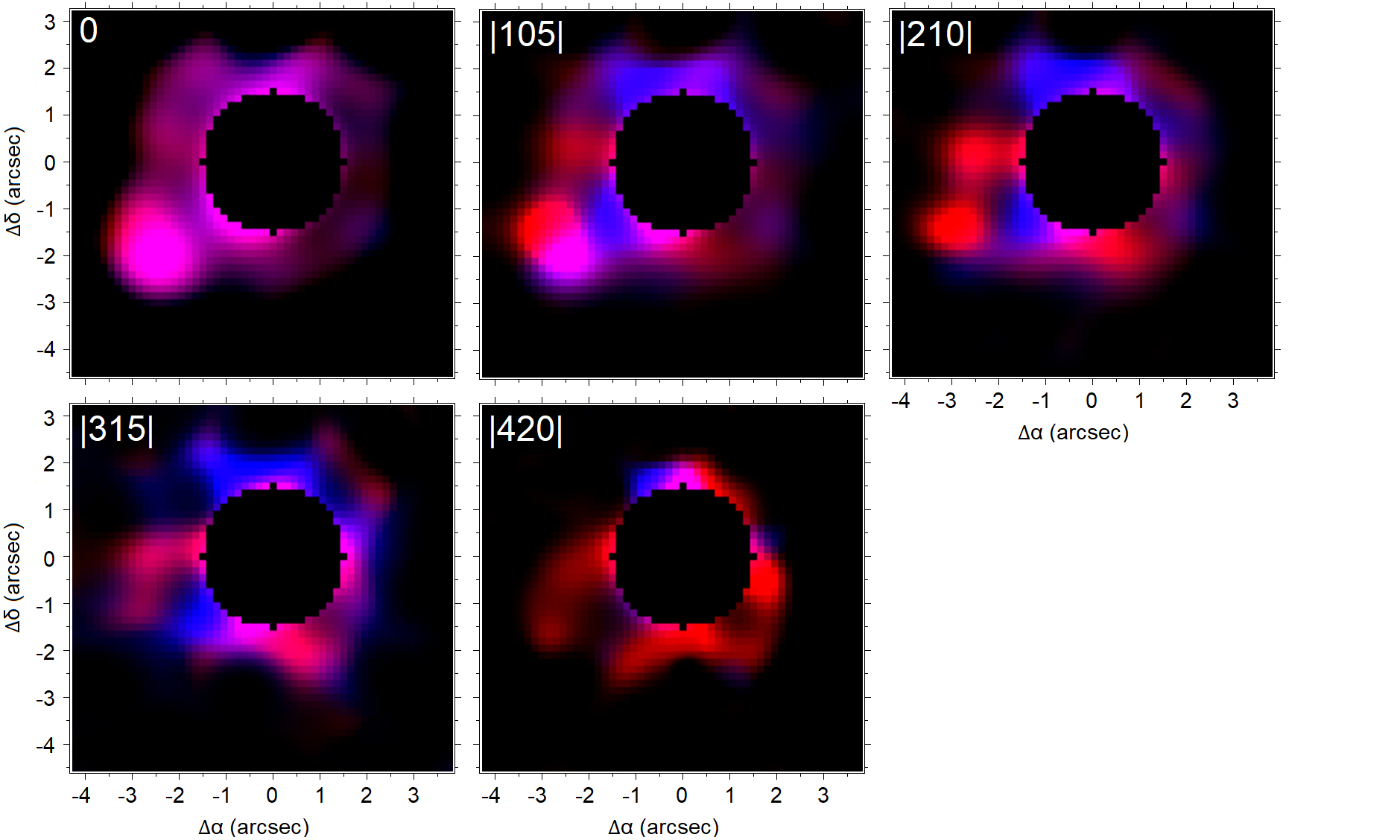}}
    \caption{BRV maps for the [\ion{O}{iii}] emission line of ESO~138-G001 derived from the IFU datacubes analysed in this work.}
    \label{fig:oiii_brv_maps}
\end{figure*}

\begin{figure*}
  \resizebox{0.90\hsize}{!}{\includegraphics{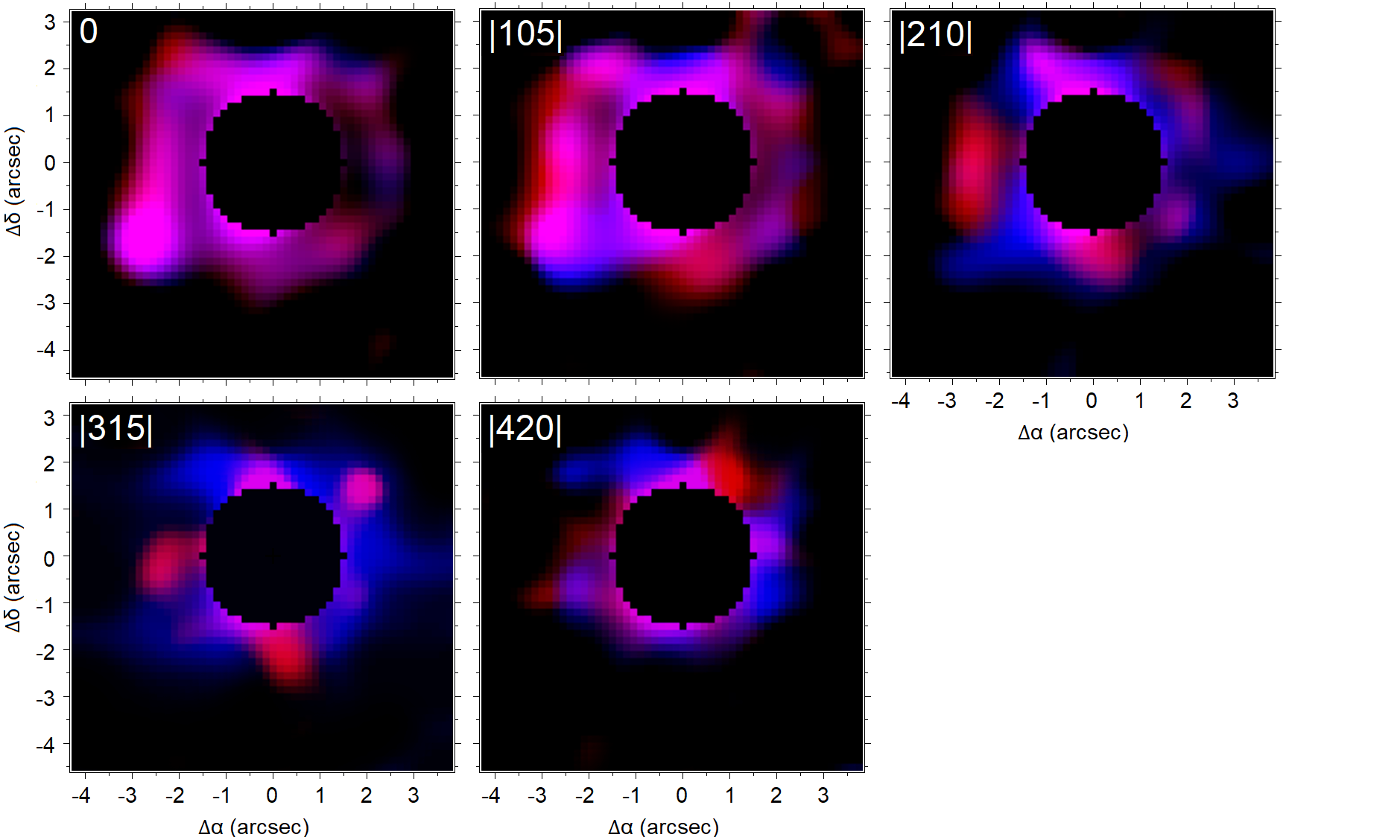}}
    \caption{Same as Figure~\ref{fig:oiii_brv_maps}  for the H$\alpha$ emission line.}
    \label{fig:ha_brv_maps}
\end{figure*}

Figures~\ref{fig:oiii_brv_maps} and~\ref{fig:ha_brv_maps} picture BRV maps or Blue and Red Velocity channel maps for [O\,{\sc iii}]~$\lambda5007$ and H$\alpha$, respectively. They combine negative and positive velocities of the same absolute value in a single map. They reveal subtleties in the velocity field for different lines. For instance, the NE portion of the blob is redder (in velocity space) in [\ion{O}{iii}] than in H$\alpha$. Moreover, the extended emission NE of the nucleus appears bluer in [\ion{O}{iii}] than in H$\alpha$. 

The BRV maps also reveal faint emissions around the nucleus, visible mostly in the high-velocity channels 315~\kms\ and 420~\kms.  We interpret them as gas located at the inner parts of the ionisation cone. In this scenario, we observe high-velocity gas connected to the nucleus and cavity-like structures with mixed kinematics with respect to the line-of-sight (LoS), possibly representing expanding shells inside the limits of the cone. The energy required for this expansion - seen here in low-velocities projected on the plane of the sky - may come from the initial stages of jet propagation clearing its way out to the nuclear region. This is thought to happen in compact nuclear radio sources where the jet is not resolved or barely elongated~\citep{morganti13b, zovaro19}.

From the values of FWHM of the emission lines (see Table~\ref{tab:fluxblobs}), we found that the line profiles in the SE blob are intrinsically narrow, lacking of a broad component, contrary to the expectations if they were in an outflow. Indeed, the FWHM points to velocities of $\sim$160~\kms, narrower than the values measured at the AGN.  

Regarding the NE extended emission, the gas at that location is significantly more perturbed kinematically than its SE counterpart. It shows up prominently in the velocity channels -420~\kms\ to -105~\kms, implying that this gas is blue-shifted from the systemic velocity. Moreover, the FWHM of the emission lines listed in Table~\ref{tab:fluxblobs} suggests that the gas is more turbulent, with values indicating velocities of $\sim$300~\kms, that is, about two times broader than the one found in the SE blob. 

The more complex gas kinematics displayed by the NE knot can be explained in terms of gas that is strongly perturbed by the passage of a radio jet. The radio jet emission in the bottom-right panel of Figure~\ref{fig:lines}  shows that the NE edge of the jet spatially coincides with the position of the NE knot. Very likely, the interaction between the jet and the ISM gas leads to an increase in gas velocity and turbulence. Similar effects have already been found in galaxies such as NGC\,4388 \citep{Ardila17}, ESO~428-G14 \citep{DMay18} and IC\,5063~\citep{Fonseca-Faria+23}. In this scenario, the passage of the nuclear radio jet inflates blobs of gas that expand adiabatically against the ISM.

\subsection{A possible nuclear outflow in ESO\,138-G001?}

From the Gaussian fitting applied to the spectra in \eso{} we detected broad emission components associated with the brightest lines in the nuclear spectrum (see Figure~\ref{fig:gauss_fit} and Table~\ref{tab:fluxnuc}). Overall, these broad features display FWHM $>$ 450~\kms, except in the [\ion{S}{ii}] lines, where they reach only $\sim$310~\kms. In [\ion{Fe}{vii}]~$\lambda$6087, the FWHM of the broad component is nearly 600~\kms. Interestingly, the centroid position in wavelength of the broad feature in the lines where it shows up is very close to that of the narrow component, with shifts of usually $<$ 1~\AA. No clear trend of the shift towards the blue or the red in the broad component relative to the narrow one is detected. We tentatively associate the broad line with an outflow. If that is the case, it is restricted to the nuclear region as lines of such FWHM are only detected in the nucleus. Moreover, the outflowing gas should be moving nearly perpendicular to the LoS of the observer due to the lack of either blueshift or redshift of the broad lines.

We searched in the literature for previous observational evidence of nuclear outflows in \eso{} but found no positive results. This is probably because most works dedicated to \eso{} have been carried out using low-resolution spectra. Here, due to our superior spectral resolution, we are able to examine in more detail the different emission lines in that object.  We rule out that the detection of the broad component can be due to the excellent S/N of the brightest lines in the nuclear spectrum. This is because we notice that it was also detected in most coronal and the [\ion{S}{ii}]~$\lambda\lambda$6716,6731 lines. These are weak features when compared to [\ion{O}{iii}]~$\lambda$5007 or H$\alpha$, where the broad component is also present. Therefore, we believe that the possibility of a nuclear outflow in \eso{} is real and first detected here in this work. Also, it is important to mention that the broad component is constrained to the central 1$\arcsec$, where the radio jet is launched. Along the path where the jet propagates, the jet compresses the ISM gas, inducing strong turbulence and launching outflows, giving rise to gas with different kinematics from that rotating with the galaxy. Another possibility is that the broad component is associated to ionised gas that is in-situ accelerated by a wind from the central source. Theoretical models \citep{Muk16} have already confirmed the viability of the jet-driven outflow scenario and successfully applied to IC\,5063 \citep{Murk18}. Evidence of the wind-driven scenario has been employed to explain the NLR of Mrk\,573 \citep{travis17}. Our data does not allow us to distinguish between these two possibilities.

\section{Theoretical predictions of the observed spectra using {\sc cloudy}}
\label{sec:cloudy-modelling}

\subsection{Modelling setup and SED}

To investigate the physical conditions in the nuclear region and in the SE blob for ESO138-G001, we performed photoionisation modelling using {\sc cloudy} \citep[version 22.01,][]{Ferland_etal_2017}. We consider separate modelling for the two regions and retrieve the intensities of the emission lines in the respective cases. For the nuclear region, we consider a grid in ionisation parameter ($U$), i.e., -4 $\leq$ log $U$ $\leq$ 1, with step size 0.25. We obtain the value for the inner radius of the cloud by assuming a power-law density law which has the following form:

\begin{equation}
n\left(r\right) = n_o\left(r_o\right)\left(\frac{r}{r_o}\right)^{\alpha}
\label{eq:density-law}
\end{equation}
here, $n_o$ is the density at the location of the onset of the cloud which is set by the electron density obtained by the [Ar {\sc iv}] emission line, i.e., 10$^{3.68}$ cm$^{-3}$. The extent of the coronal line emission is estimated using the FWHM of the [Fe\,{\sc vii}] emitting region (see Figure \ref{fig:lines}), i.e., 0.8 arcsec, which is approximately 146 pc. We take half of this size as the outer radius of the emitting medium, which is roughly equal to 73 pc (= $r$). Assuming the density law with $\alpha$ = -1 and the density at the outer edge of the cloud set by the one determined from the [S {\sc ii}] emission lines, i.e., 10$^{2.75}$ cm$^{-3}$, we obtain the location of the inner radius to be 8.9 pc (or $\approx$ 
2.75$\times 10^{19}$ cm). We also consider an alternate density law with an $\alpha$ = -2. This sets the inner radius much further out, at 26 pc ($\approx$ 8$\times 10^{19}$ cm).

Since we have the extension of the cloud ($d$ = $r$ - $r_o$), we have the stopping criterion\footnote{{\sc cloudy} allows multiple ways to impose the stopping criteria, e.g., setting a minimum temperature, or a minimum optical depth, etc. The radiative transfer computations are stopped when the prescribed value is reached. In our case, we utilise the stop column density command to execute this.} for our simulation set by the cloud column density. We integrate the Eq. \ref{eq:density-law} for the two cases of $\alpha$, which gives us the following equations for the column density:

\begin{equation}
\begin{split}
\alpha = -1 \rightarrow & N_{H} = n_o\left(r_o\right)r_o{\rm ln}\left(\frac{r}{r_o}\right) \\
\alpha = -2 \rightarrow & N_{H} = n_o\left(r_o\right)r_o\left(\frac{r - r_o}{r}\right)
\end{split}
\end{equation}

This value is 10$^{23.08}$ cm$^{-2}$ and 10$^{23.4}$ cm$^{-2}$ for $\alpha$ = -1 and -2, respectively. Additionally, we assume solar abundances for the ionised clouds in our modelling.   

\begin{figure}
    \centering
    \includegraphics[width=\columnwidth]{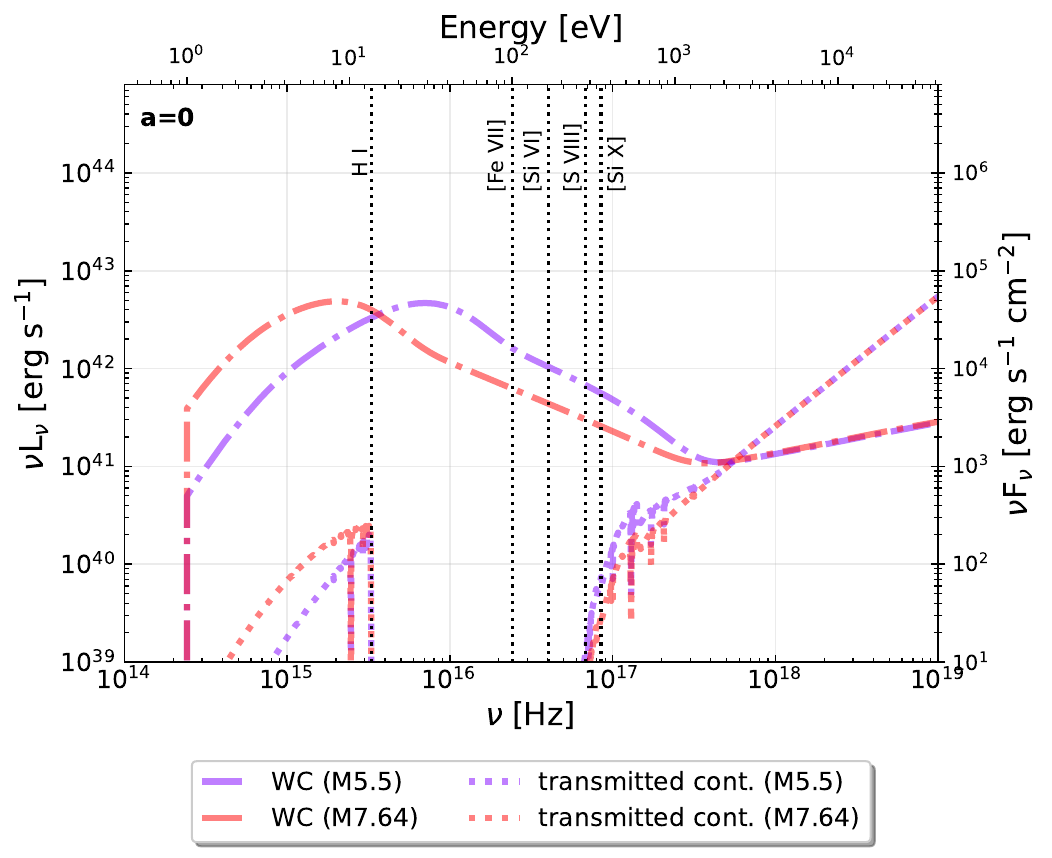}
    \caption{SEDs assumed in our {\sc cloudy} modelling. The incident continua for the two black hole mass cases are shown in dot-dashed. These are used for the modelling of the nuclear region. The corresponding transmitted continua from these modelling are shown in dotted lines. These latter distributions are used as incident continua for the SE blob. These SEDs have been made assuming a black hole spin parameter, a = 0.}
    \label{fig:sed}
\end{figure}

For the ionising continuum, we model the broad-band spectral energy distribution (from 1 eV up to a few times 10 keV) using the {\sc xspec} model {\sc AGNSED} \citep{Kubota_Done_18}. We assume two cases of black hole mass: (a) 10$^{5.50}$~M$_{\odot}$ from X-ray variability using XMM-Newton 0.5-10 keV spectrum \citep{Hernandez-Garcia_etal_2015}; and (b) 10$^{7.64}$~M$_{\odot}$ \citep{Cerqueira-Campos_etal_2021}. The corresponding SED is shown in Figure~\ref{fig:sed}. We assume a constant value for the Eddington ratio ($\lambda_{Edd}$ = 0.01, \citealt{Hernandez-Garcia_etal_2015}) as there is no estimate provided in \citet{Cerqueira-Campos_etal_2021}. However, we note that the increased black hole mass at the same Eddington ratio corresponds to a lowering in the accretion disk temperature by $\sim$3.42 \citep[see e.g.,][]{Panda_etal_2017, Panda_etal_2018}, i.e., the accretion disk temperature is lowered by this factor. This eventually affects the bolometric luminosity (or the area under the curve as shown in Figure \ref{fig:sed}). In addition, we assume the spin parameter, a = 0, and the redshift, z = 0.009140, keeping all the remaining parameters of the {\sc AGNSED} model at their default values. In Figure \ref{fig:sed}, we mark the location of some of the prominent coronal emission lines and the hydrogen ionisation front at 13.6 eV (= 1 Rydberg).

\subsection{Spectral synthesis of the nuclear region and the SE blob}

A preliminary result from the modelling is that the change in the density from the inner to the outer edge of the cloud is small (we notice a minor drop in density within the extent of the cloud, i.e., the density contrast is about $\sim$8.5, estimated for the densities at the inner (10$^{3.68}$ cm$^{-3}$) and outer (10$^{2.75}$ cm$^{-3}$) radii, respectively). Assuming a constant density model in our setup does not affect much the predicted line ratios. Therefore, earlier approaches of assuming constant density models are equally acceptable \citep[see e.g.,][]{Prieto_etal_2022} in this case. Even so, in this work, we continue with our original assumption of the density law.

\begin{figure*}
    \centering
    \includegraphics[width=0.495\textwidth]{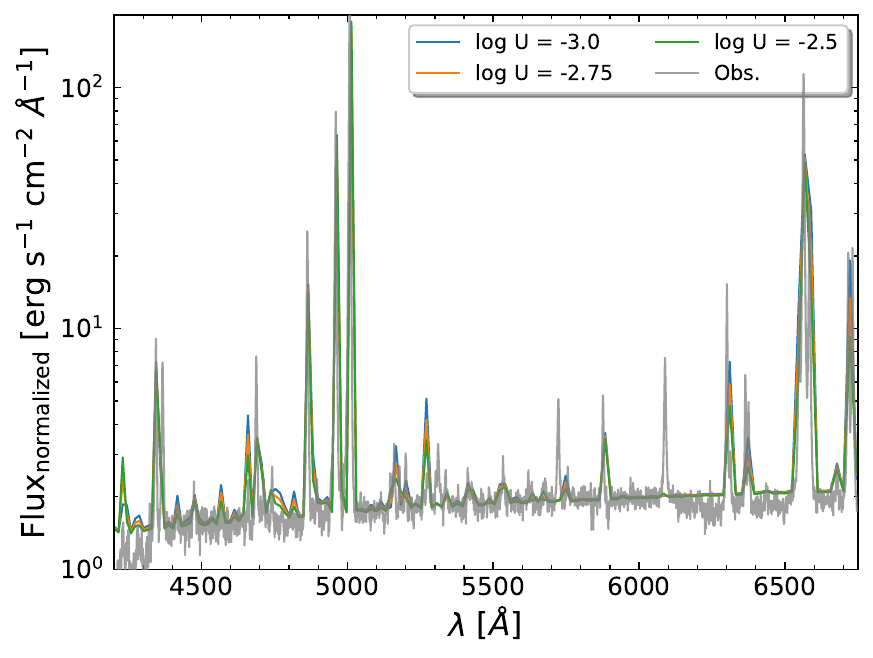}
    \includegraphics[width=0.495\textwidth]{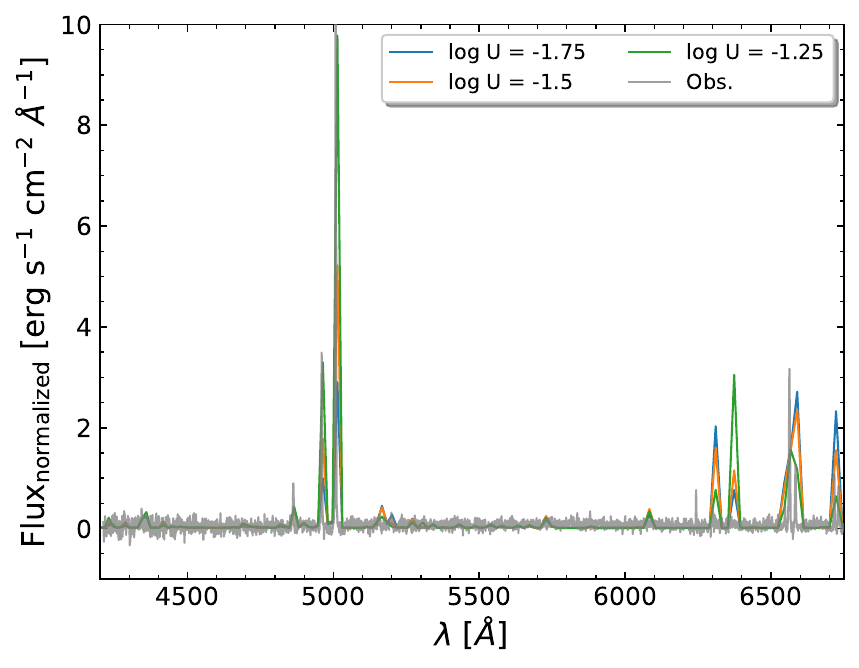}
    \caption{Synthetic spectra comparison with observed spectrum (in grey), for the nuclear region (left panel), and the SE blob (right panel). The modelled line dispersion is set to the minimum limit by the model emission line grid. The values of the ionisation parameter that best represents the observed spectrum in each case are noted in the legend in each panel. We note here that the incident continua used in the modelling of these two regions correspond to the dot-dashed SED (for the nuclear region), and the dotted SED (for the SE blob) for the black hole mass of 10$^{\rm 5.50}$ M$_{\odot}$ (shown with purple colour in Figure \ref{fig:sed}).}
    \label{fig:compare-spectrum}
\end{figure*}

We extract the synthetic spectrum (with emission lines and continuum) predicted by {\sc cloudy} from each of these models. In the left panel of Figure \ref{fig:compare-spectrum}, we show representative modelled spectra (corresponding to a given ionisation parameter) and compare them against the observed spectrum from the nuclear region (in grey). The modelled spectrum for an ionisation parameter, -2.75 $\lesssim$ log U $\lesssim$ -2.25, is able to reproduce the intensities of most of the strong emission features in the observed spectrum. The modelled spectrum within this range of ionization parameters gives the lowest residual when compared against the observed spectrum from the nuclear region.

\begin{figure*}
    \centering
    \includegraphics[width=\textwidth]{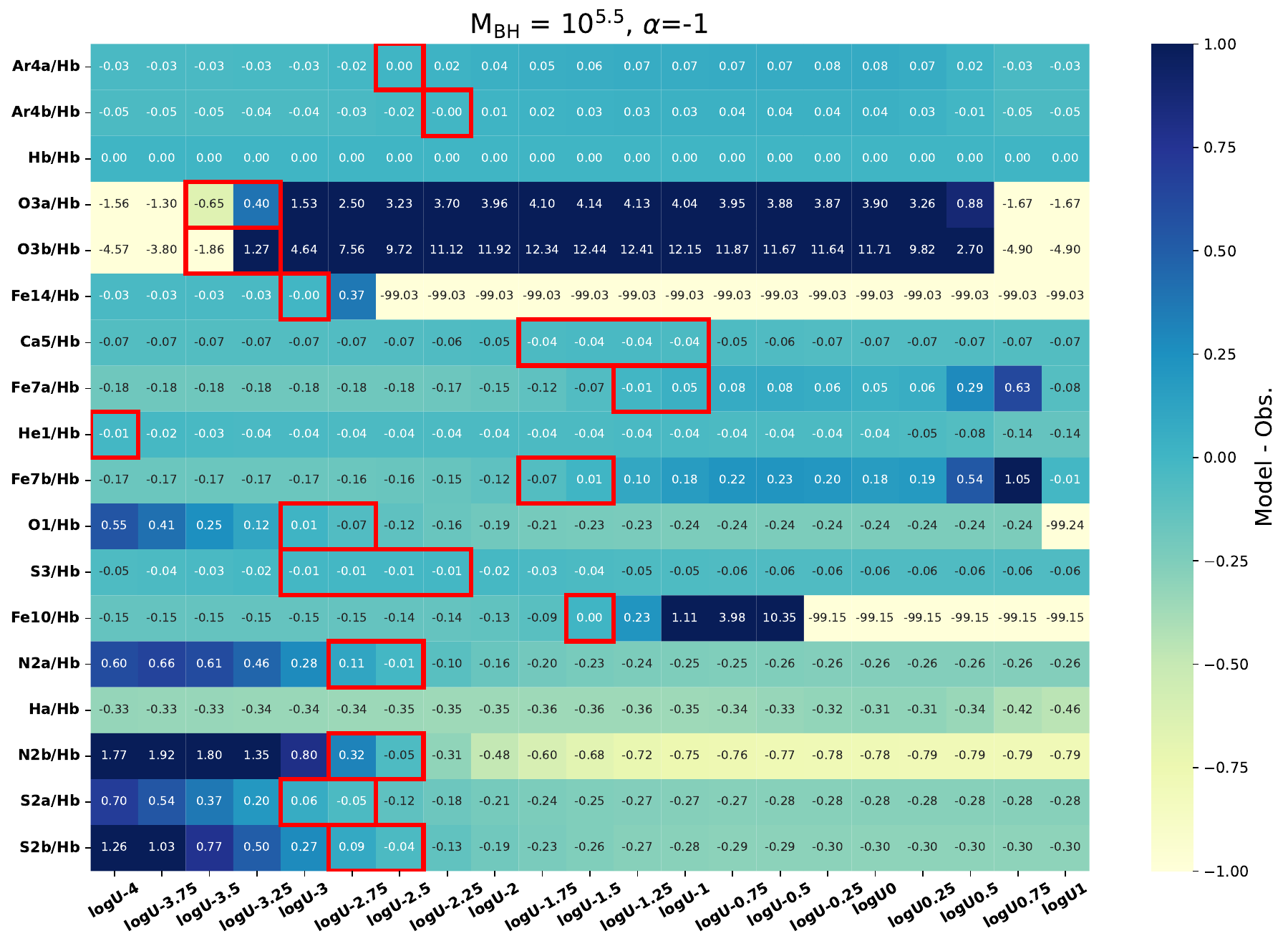}
    \caption{Heatmaps for the BH mass case, M$_{\rm BH}$ = 10$^{5.50}$ M$_{\odot}$ for the density law slope, $\alpha$ = -1, for the nuclear region. Each of the considered emission lines for the outflowing component is normalised to the H$\beta$ line. The x-axis represents the range of ionisation parameters considered in our models in each panel. The cases of ionisation parameters per each line ratio that have the smallest residuals (Modelled ratio - Observed ratio) are highlighted with red boxes.}
    \label{fig:heatmap-1}
\end{figure*}

We take the modelling aspect a step further to highlight the ideal location of a given emission line based on its strength with respect to H$\beta$, i.e., we estimate the range of ionization parameters required to emit the given line. The value of the ionization parameter coupled with the value of the density of the emitting medium and for a given SED allows for recovery of the radial location of the medium itself \citep{Wandel1999, Negrete2014, Panda_2021, Panda2022}. From the observational side, we consider the prominent emission lines from the outflowing component of the nuclear region in the observed spectrum, tabulated in Table \ref{tab:fluxnuc}. We then take the ratio of the line flux for the said emission line to the H$\beta$ flux. We perform an identical procedure to obtain the modelled flux ratios. In Figure \ref{fig:heatmap-1}, we show a heatmap distribution between the prominent line flux ratios versus the various models considered in this work (in terms of the ionisation parameter). The colour scale indicates the residual between the modelled flux ratio to the observed flux ratio per given emission line. We further highlight the location of the minimum (or minima where the residuals change signs) using rectangular regions (in red). This allows us to infer the corresponding value (or range of values) for the ionisation parameter required to recover the various flux ratios.

We notice that the emission lines with lower ionisation states (e.g., [O {\sc iii}], [N {\sc ii}] and [S {\sc ii}]) require very low ionisation parameters (or photon flux), i.e., as low as log U $\lesssim$ -3.75, to recover the observed intensities, while coronal emission lines (e.g., [Fe {\sc vii}] and [Ca {\sc v}]) are produced in optimal quantity when the ionisation parameter is larger by at least 1-2~dex \footnote{We highlight one such case in Figure \ref{fig:cloudy_fe7} where the model with log U = -1 reproduces the observed [Fe {\sc vii}] emission with the lowest residual. Although this particular model doesn't optimise the emission for some of the other CLs in the observed range.}. Higher ionisation states of Fe-based coronal lines ([Fe {\sc x}] and [Fe {\sc xiv}]) are also recovered, although the observed strength of these lines is fairly low. We, therefore, do not emphasise the recovery of these lines since they are suggested to be produced much closer to the central nuclei, where the ionisation is stronger to produce lines of such high ionisation states. We show the remaining heatmaps in Figures \ref{fig:heatmap-alt1}, \ref{fig:heatmap-alt2} and \ref{fig:heatmap-alt3}. 

We note that the range of suitable ionization parameters for the high and the low ionization lines for the two cases of $\alpha$ for a given black hole mass rather stays constant. Also, an increase in the BH mass (from 10$^{5.50}$ to 10$^{7.64}$ M$_{\odot}$) does not lead to a marked difference in the heatmaps. We note here that we have assumed the same accretion rate while constructing the two SEDs, i.e., $\lambda_{Edd}$ = 0.01. This value of the Eddington ratio is obtained from \citet{Hernandez-Garcia_etal_2015}, which is the only Eddington ratio estimate available for this source. At a higher black hole mass, this value of the Eddington ratio may be lower. Hence the two SEDs that are used in this work, give similar results for the line emission. The only visible difference is seen in the recovery of the permitted lines, e.g., He {\sc i}/H$\beta$, where, in the case of the higher black hole mass, the required ionization parameter range shows two groups of solutions - one at the lowest end (log U $\lesssim -4$) and another which extends to higher values for the ionization parameter, i.e., -1 $\lesssim$ log U $\lesssim$ 0.75. This range is restricted only to the first group of solutions in the case of the lower black hole mass case. In addition to this, considering the two cases of density law does not show a marked difference in the recovery of the emission line ratios, giving similar values for the ionization parameters range per emission line ratio. We also note that the assumption of a $\alpha$-dependent stopping criterion, i.e., cloud column density (N$_{\rm H}$), relative to using a constant value, e.g., 10$^{24}$ cm$^{-2}$ give almost identical ionization parameters for the line ratios considered. The value, 10$^{24}$ cm$^{-2}$, is obtained assuming N$_{\rm H}$ = $d$*$n_o$, where $d = r - r_o$ and $n_o$ = 10$^{3.68}$ cm$^{-3}$. We find only minor shifts for [Fe{\sc xiv}]/H$\beta$, where for both $\alpha$ values and BH mass = 10$^{5.50}$ M$_{\odot}$, the log U shift by +0.25 dex for the constant column density case. This is found to be true also for $\alpha$ = -1 and BH mass = 10$^{7.64}$ M$_{\odot}$. In this latter case, we also notice that the log U range extends to higher values for He {\sc i}/H$\beta$, while the lowest residuals for [Fe{\sc vii}]/H$\beta$ are recovered for slightly lower log U values. For $\alpha$ = -2 and BH mass = 10$^{7.64}$ M$_{\odot}$, we do not see any change between the constant and $\alpha$-dependent column density.

We made a similar analysis for the SE blob but assume a density, of 10$^{2.09}$ cm$^{-3}$ (see Table \ref{tab:properties}), for the outer region of the cloud. In this case, we utilise the transmitted continuum that comes out of the cloud emitting the lines in the nuclear region as described earlier. This results in a diluted continuum received by the blob (see the dotted curves in Figure \ref{fig:sed}). We notice that the best-fit ionisation parameter lies in the range of -1.75 $\lesssim$ log $U \lesssim$ -1.25, in order to recover the intensity of the prominent emission lines for the SE blob (see the right panel of Figure \ref{fig:compare-spectrum}). The observed spectrum is shown in grey while the modelled spectra for the three cases of ionisation parameters (log $U$ = -1.75, -1.5 and -1.25) are shown in blue, orange and green, respectively. This ionisation parameter range corresponds to the diluted continuum - which is $\sim$4\% of the original incident continuum (i.e., the bolometric output based on the SED) originating from the accretion disk (this is equivalent to the ratio of the area of the dot-dashed curve and its corresponding transmitted continuum curve at about the 1 Rydberg limit for either of the black hole mass case). It is important to mention that if one uses the original incident continuum and scales the ionisation with the location of the SE blob and its density, this does not give the same resulting spectrum. In this latter case, the spectrum shows coronal emission lines with relatively high intensities, which is contrary to what the observed spectrum of the SE blob shows. In our approach, we first pass the incident continuum through a thick slab of gas (one that is responsible for producing the lines from the nuclear region) and the transmitted continuum that comes out of this cloud is the one responsible for the line production in the SE blob. The transmitted continuum (see the dotted curves in Figure \ref{fig:sed}) almost becomes negligible at the ionisation energies of the prominent coronal lines. As a result, the high ionisation forbidden lines are almost non-existent in our modelled spectrum and hardly any emission from the prominent coronal lines in this wavelength range is recovered from this modelling - which is in agreement with the observed spectrum. In this way, we are successful in recovering the physical conditions of the SE blob and its spectrum - which only shows forbidden lines of low ionisation states, e.g., [O {\sc iii}] and [S {\sc ii}], in addition to the narrow Balmer lines. The results obtained for the NE knot are almost identical to the results obtained for the SE blob, requiring -2 $\lesssim$ log U $\lesssim$ -1.5 to reproduce the observed emission spectrum.


\section{Discussion and Final Remarks}
\label{sec:concl}

Two main aspects of \eso{} are investigated in this paper. The first one is related to the origin, morphology and size of the prominent coronal line region (CLR) displayed by this AGN. The second one is about the nature of the SE emission blob previously reported in the literature \citep{Ferruit00}. The data studied here by means of SIFS/SOAR spectroscopy allow us to set important constraints on these two matters.

Our results point out that the coronal line region in \eso{} is compact and barely resolved under the angular scale of $\sim$0.62 arcsec of our data. Both [Fe\,{\sc vii}]~6087~\AA\ and [Fe\,{\sc x}]~$\lambda6374$ are contained within the inner 0.8$\arcsec$ (FWHM). The former ion displays a hint of extended emission to the SE. Neither the SE blob nor the NE knot shows hints of coronal emission. We, therefore, conclude that the bulk of the CLR in \eso{} is restricted to a circular region 146~pc in diameter, centred at the AGN.  
  
The above result strongly contrasts with other nearby AGNs, also known for displaying a bright CLR such as Circinus or IC\,5063. In the latter two sources, the CLR is extended to several hundred parsecs or even a few thousand parsecs away from the central engine \citep{Rodriguez-Ardila_Fonseca-Faria20, Fonseca-Faria+23}.   

The above findings are in agreement with the ones reported in the X-rays. \citet{DeCicco15} concluded that the X-ray emitting gas is \eso{} is rather compact, supporting our findings of a very compact CLR, likely also produced by radiation of the AGN. Sources known for displaying an extended CLR at the scale of several hundreds of parsecs (i.e., NGC\,4388, Circinus, among others) usually show extended X-ray emission at the same scale \citep{matt94,smith01,iwasawa03}.  Moreover, \citet{DeCicco15} found that \eso{} appears unresolved although hints of an elongation along the NW-SE direction, where the SE blob is located, are observed. The faintness of the X-ray emission at the blob location could be explained by considering that the X-ray flux of the NLR is generally a factor 2.8$-$11 lower than the [O\,{\sc iii}] flux \citep{Bianchi06}. From our detected flux of [O\,{\sc iii}]~$\lambda5007$ (see Table~\ref{tab:fluxblobs}), the X-ray would be probably undetected.

The deblending procedure applied to the emission lines in the nuclear spectrum revealed the presence of broad components associated with the most relevant lines. That component displays FWHM of $\sim$450~\kms\ and reaches up to 600~\kms\ in the coronal lines. We associated this broad characteristic with a nuclear outflow, first reported in this work, and mostly photoionised by radiation from the central engine. 

{\sc cloudy} modelling of the observed nuclear spectrum confirms our claims of a very compact CLR, powered by photoionisation from the AGN. Contribution from shocks seems very unlikely, being its major effect in perturbing kinematically the gas but unable to drive shocks to excite coronal lines.  

Regarding the nature of the SE blob, all the evidence points out that it consists of NLR gas that is photoionised by the modified AGN continuum that is filtered by gas clouds located in the inner tens of parsecs of the central source. From the grid of models using realistic SED, we found out that the blob cannot be directly illuminated by the AGN continuum, otherwise, emission lines from higher ionisation states would show up. The fact that the NE knot also lacks emission lines of ionization potential  $>$ 35~eV  suggests that the central region filters out the high-energy photons from the AGN. It should be noticed, though, that evidence of an extended NLR  (ENLR) in this galaxy was provided by \citet{Schmitt_Storchi-Bergmann95}, who traced the ENLR over about 12$\arcsec$, out to the inner jet region. Moreover, in the WiFeS data cube, \citet{dopita+15} reports an ENLR
extending over the full field, out to at least 35 arcsecs (6.4 kpc) in diameter.  Within the ENLR, the highest ionisation line detected is [\ion{O}{iii}], in agreement with the results shown here. We rule out star formation as the main excitation mechanism due to the lack of a blue continuum and stellar absorption features due to massive, hot stars.

\begin{figure}
    \resizebox{0.90\hsize}{!}{\includegraphics{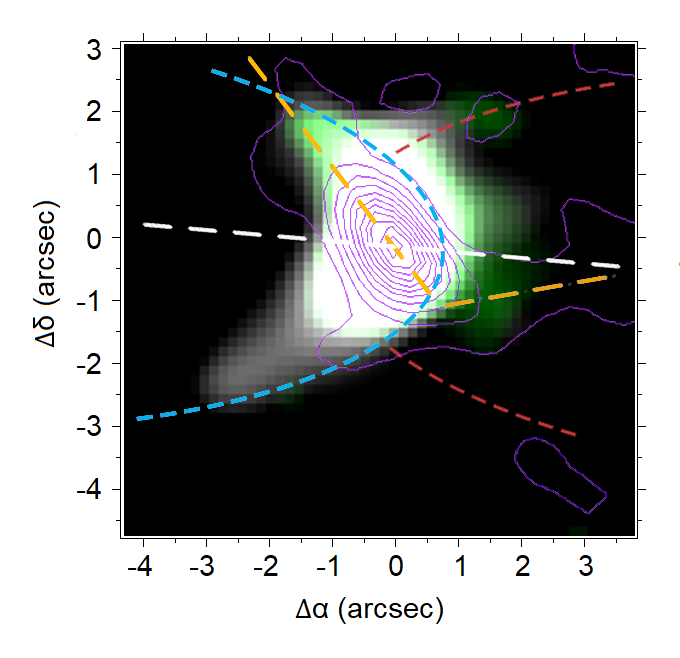}}
    \caption{Blue-shifted emissions of [\ion{O}{iii}], centred in -210 km\,s$^{-1}$~(grey), and [\ion{S}{ii}] centred in -180 km\,s$^{-1}$~(green), chosen to better highlight the ionisation cone and counter cone edges (dashed blue and dashed red curves, respectively). The white dashed line represents the axis of the ionisation cone, found in this work. The contours in purple show the radio emission and the jet direction \citep[yellow dashed lines,][]{morganti99}. North is up and East is to the left. }
     \label{fig:sii_g_oiii_g_blueshift_cone_scale}
    \end{figure}

From the evidence gathered throughout this work, we propose a scenario depicted in Figure~\ref{fig:sii_g_oiii_g_blueshift_cone_scale} for the central region of \eso{}. Basically, the SE blob and the NE knot correspond to the edges of the ionisation cone, represented in Figure~\ref{fig:sii_g_oiii_g_blueshift_cone_scale} by the dashed blue line. The radio jet is propagating at a PA almost grazing the northern edge of the cone. The interaction of the jet with the nuclear gas would produce an enhancement of the gas turbulence, reflected in the FWHM of the emission lines, and very likely responsible for a putative nuclear outflow, first detected in this work. The SE blob corresponds to the southern edge of the cone, very close to the AGN. The gas is moving nearly perpendicular to the LoS so that no significant blueshift in the centroid position of the emission lines is detected. Moreover, because of the little influence of the jet on the southern edge of the cone, the gas kinematics is little affected. The counter-cone, represented by the dashed red line, is not well-defined likely because of the obscuration due to the host galaxy. In Figure~\ref{fig:sii_g_oiii_g_blueshift_cone_scale}, the dashed white line marks the PA of the cone. This structure is extended further out, to at least 35$\arcsec$ (6.4~kpc) in diameter, according to \citet{dopita+15}. Interestingly, the SED that illuminates the clouds in the ionisation cone is filtered out by the central emission, so that only low- and mid-ionisation lines are detected. That would explain the compactness of the CLR in that source.

\section*{Acknowledgements}

We thank the anonymous referee for useful comments and suggestions that helped to improve this manuscript. ARA acknowledges Conselho Nacional de Desenvolvimento Cient\'{\i}fico e Tecnol\'ogico (CNPq) for partial support to this work through grant 312036/2019-1. SP acknowledges the Conselho Nacional de Desenvolvimento Científico e Tecnológico (CNPq) Fellowship 300936/2023-0. LF acknowledges the partial financial support from FAPEMIG (Projeto Universal, APQ-02423-21) and from CNPq (Projeto Universal 436269/2018-0).  

\section*{Data Availability}

 Observed and simulated data will be made available upon reasonable request to the authors.



\bibliographystyle{mnras}
\bibliography{ms} 

\begin{thebibliography}{}
\makeatletter
\relax
\def\mn@urlcharsother{\let\do\@makeother \do\$\do\&\do\#\do\^\do\_\do\%\do\~}
\def\mn@doi{\begingroup\mn@urlcharsother \@ifnextchar [ {\mn@doi@}
  {\mn@doi@[]}}
\def\mn@doi@[#1]#2{\def\@tempa{#1}\ifx\@tempa\@empty \href
  {http://dx.doi.org/#2} {doi:#2}\else \href {http://dx.doi.org/#2} {#1}\fi
  \endgroup}
\def\mn@eprint#1#2{\mn@eprint@#1:#2::\@nil}
\def\mn@eprint@arXiv#1{\href {http://arxiv.org/abs/#1} {{\tt arXiv:#1}}}
\def\mn@eprint@dblp#1{\href {http://dblp.uni-trier.de/rec/bibtex/#1.xml}
  {dblp:#1}}
\def\mn@eprint@#1:#2:#3:#4\@nil{\def\@tempa {#1}\def\@tempb {#2}\def\@tempc
  {#3}\ifx \@tempc \@empty \let \@tempc \@tempb \let \@tempb \@tempa \fi \ifx
  \@tempb \@empty \def\@tempb {arXiv}\fi \@ifundefined
  {mn@eprint@\@tempb}{\@tempb:\@tempc}{\expandafter \expandafter \csname
  mn@eprint@\@tempb\endcsname \expandafter{\@tempc}}}

\bibitem[\protect\citeauthoryear{{Alloin}, {Bica}, {Bonatto}  \&
  {Prugniel}}{{Alloin} et~al.}{1992}]{Alloin92}
{Alloin} D.,  {Bica} E.,  {Bonatto} C.,   {Prugniel} P.,  1992, \aap, \href
  {https://ui.adsabs.harvard.edu/abs/1992A&A...266..117A} {266, 117}

\bibitem[\protect\citeauthoryear{{Astropy Collaboration} et~al.,}{{Astropy
  Collaboration} et~al.}{2013}]{astropy:2013}
{Astropy Collaboration} et~al., 2013, \mn@doi [\aap]
  {10.1051/0004-6361/201322068}, \href
  {http://adsabs.harvard.edu/abs/2013A%26A...558A..33A} {558, A33}

\bibitem[\protect\citeauthoryear{{Astropy Collaboration} et~al.,}{{Astropy
  Collaboration} et~al.}{2018}]{astropy:2018}
{Astropy Collaboration} et~al., 2018, \mn@doi [\aj] {10.3847/1538-3881/aabc4f},
  \href {https://ui.adsabs.harvard.edu/abs/2018AJ....156..123A} {156, 123}

\bibitem[\protect\citeauthoryear{{Astropy Collaboration} et~al.,}{{Astropy
  Collaboration} et~al.}{2022}]{astropy:2022}
{Astropy Collaboration} et~al., 2022, \mn@doi [apj] {10.3847/1538-4357/ac7c74},
  \href {https://ui.adsabs.harvard.edu/abs/2022ApJ...935..167A} {935, 167}

\bibitem[\protect\citeauthoryear{{Bianchi}, {Guainazzi}  \&
  {Chiaberge}}{{Bianchi} et~al.}{2006}]{Bianchi06}
{Bianchi} S.,  {Guainazzi} M.,   {Chiaberge} M.,  2006, \mn@doi [\aap]
  {10.1051/0004-6361:20054091}, \href
  {https://ui.adsabs.harvard.edu/abs/2006A&A...448..499B} {448, 499}

\bibitem[\protect\citeauthoryear{{Bianchin} et~al.,}{{Bianchin}
  et~al.}{2022}]{Bianchin22}
{Bianchin} M.,  et~al., 2022, \mn@doi [\mnras] {10.1093/mnras/stab3468}, \href
  {https://ui.adsabs.harvard.edu/abs/2022MNRAS.510..639B} {510, 639}

\bibitem[\protect\citeauthoryear{{Cardelli}, {Clayton}  \& {Mathis}}{{Cardelli}
  et~al.}{1989}]{ccm89}
{Cardelli} J.~A.,  {Clayton} G.~C.,   {Mathis} J.~S.,  1989, \mn@doi [\apj]
  {10.1086/167900}, \href
  {https://ui.adsabs.harvard.edu/abs/1989ApJ...345..245C} {345, 245}

\bibitem[\protect\citeauthoryear{{Cerqueira-Campos}, {Rodr{\'\i}guez-Ardila},
  {Riffel}, {Marinello}, {Prieto}  \& {Dahmer-Hahn}}{{Cerqueira-Campos}
  et~al.}{2021}]{Cerqueira-Campos_etal_2021}
{Cerqueira-Campos} F.~C.,  {Rodr{\'\i}guez-Ardila} A.,  {Riffel} R.,
  {Marinello} M.,  {Prieto} A.,   {Dahmer-Hahn} L.~G.,  2021, \mn@doi [\mnras]
  {10.1093/mnras/staa3320}, \href
  {https://ui.adsabs.harvard.edu/abs/2021MNRAS.500.2666C} {500, 2666}

\bibitem[\protect\citeauthoryear{{Collinge} \& {Brandt}}{{Collinge} \&
  {Brandt}}{2000}]{Collinge_Brandt00}
{Collinge} M.~J.,  {Brandt} W.~N.,  2000, \mn@doi [\mnras]
  {10.1046/j.1365-8711.2000.03871.x}, \href
  {https://ui.adsabs.harvard.edu/abs/2000MNRAS.317L..35C} {317, L35}

\bibitem[\protect\citeauthoryear{Craig et~al.,}{Craig et~al.}{2017}]{ccdproc}
Craig M.,  et~al., 2017, astropy/ccdproc: v1.3.0.post1,
  \mn@doi{10.5281/zenodo.1069648}, \url
  {https://doi.org/10.5281/zenodo.1069648}

\bibitem[\protect\citeauthoryear{{De Cicco}, {Marinucci}, {Bianchi},
  {Piconcelli}, {Violino}, {Vignali}  \& {Nicastro}}{{De Cicco}
  et~al.}{2015}]{DeCicco15}
{De Cicco} M.,  {Marinucci} A.,  {Bianchi} S.,  {Piconcelli} E.,  {Violino} G.,
   {Vignali} C.,   {Nicastro} F.,  2015, \mn@doi [\mnras]
  {10.1093/mnras/stv1702}, \href
  {https://ui.adsabs.harvard.edu/abs/2015MNRAS.453.2155D} {453, 2155}

\bibitem[\protect\citeauthoryear{{Dopita} et~al.,}{{Dopita}
  et~al.}{2015}]{dopita+15}
{Dopita} M.~A.,  et~al., 2015, \mn@doi [\apjs] {10.1088/0067-0049/217/1/12},
  \href {https://ui.adsabs.harvard.edu/abs/2015ApJS..217...12D} {217, 12}

\bibitem[\protect\citeauthoryear{{Emonts}, {Morganti}, {Tadhunter},
  {Oosterloo}, {Holt}  \& {van der Hulst}}{{Emonts} et~al.}{2005}]{Emonts05}
{Emonts} B.~H.~C.,  {Morganti} R.,  {Tadhunter} C.~N.,  {Oosterloo} T.~A.,
  {Holt} J.,   {van der Hulst} J.~M.,  2005, \mn@doi [\mnras]
  {10.1111/j.1365-2966.2005.09354.x}, \href
  {https://ui.adsabs.harvard.edu/abs/2005MNRAS.362..931E} {362, 931}

\bibitem[\protect\citeauthoryear{{Faes}, {Tokovinin}, {Vieira}, {Mello},
  {Domingues}, {Andrade}, {Quint}  \& {dos Santos}}{{Faes}
  et~al.}{2018}]{Moser18}
{Faes} D.~M.,  {Tokovinin} A.,  {Vieira} T.,  {Mello} A.,  {Domingues} M.,
  {Andrade} D.,  {Quint} B.~C.,   {dos Santos} J.~B.,  2018, in {Close} L.~M.,
  {Schreiber} L.,   {Schmidt} D.,  eds,  Society of Photo-Optical
  Instrumentation Engineers (SPIE) Conference Series Vol. 10703, Adaptive
  Optics Systems VI. p. 107033C (\mn@eprint {arXiv} {1806.06742}),
  \mn@doi{10.1117/12.2312205}

\bibitem[\protect\citeauthoryear{{Ferland} et~al.,}{{Ferland}
  et~al.}{2017}]{Ferland_etal_2017}
{Ferland} G.~J.,  et~al., 2017, \mn@doi [\rmxaa] {10.48550/arXiv.1705.10877},
  \href {https://ui.adsabs.harvard.edu/abs/2017RMxAA..53..385F} {53, 385}

\bibitem[\protect\citeauthoryear{{Ferruit}, {Wilson}  \& {Mulchaey}}{{Ferruit}
  et~al.}{2000}]{Ferruit00}
{Ferruit} P.,  {Wilson} A.~S.,   {Mulchaey} J.,  2000, \mn@doi [\apjs]
  {10.1086/313379}, \href
  {https://ui.adsabs.harvard.edu/abs/2000ApJS..128..139F} {128, 139}

\bibitem[\protect\citeauthoryear{{Fischer} et~al.,}{{Fischer}
  et~al.}{2017}]{travis17}
{Fischer} T.~C.,  et~al., 2017, \mn@doi [\apj] {10.3847/1538-4357/834/1/30},
  \href {https://ui.adsabs.harvard.edu/abs/2017ApJ...834...30F} {834, 30}

\bibitem[\protect\citeauthoryear{{Fonseca-Faria}, {Rodr{\'\i}guez-Ardila},
  {Contini}, {Dahmer-Hahn}  \& {Morganti}}{{Fonseca-Faria}
  et~al.}{2023}]{Fonseca-Faria+23}
{Fonseca-Faria} M.~A.,  {Rodr{\'\i}guez-Ardila} A.,  {Contini} M.,
  {Dahmer-Hahn} L.~G.,   {Morganti} R.,  2023, \mn@doi [arXiv e-prints]
  {10.48550/arXiv.2306.09570}, \href
  {https://ui.adsabs.harvard.edu/abs/2023arXiv230609570F} {p. arXiv:2306.09570}

\bibitem[\protect\citeauthoryear{{Hern{\'a}ndez-Garc{\'\i}a}, {Masegosa},
  {Gonz{\'a}lez-Mart{\'\i}n}  \& {M{\'a}rquez}}{{Hern{\'a}ndez-Garc{\'\i}a}
  et~al.}{2015}]{Hernandez-Garcia_etal_2015}
{Hern{\'a}ndez-Garc{\'\i}a} L.,  {Masegosa} J.,  {Gonz{\'a}lez-Mart{\'\i}n} O.,
    {M{\'a}rquez} I.,  2015, \mn@doi [\aap] {10.1051/0004-6361/201526127},
  \href {https://ui.adsabs.harvard.edu/abs/2015A&A...579A..90H} {579, A90}

\bibitem[\protect\citeauthoryear{{Husemann}, {Kamann}, {Sandin}, {S{\'a}nchez},
  {Garc{\'\i}a-Benito}  \& {Mast}}{{Husemann} et~al.}{2012}]{Husemann2012}
{Husemann} B.,  {Kamann} S.,  {Sandin} C.,  {S{\'a}nchez} S.~F.,
  {Garc{\'\i}a-Benito} R.,   {Mast} D.,  2012, \mn@doi [\aap]
  {10.1051/0004-6361/201220102}, \href
  {https://ui.adsabs.harvard.edu/abs/2012A&A...545A.137H} {545, A137}

\bibitem[\protect\citeauthoryear{{Iwasawa}, {Wilson}, {Fabian}  \&
  {Young}}{{Iwasawa} et~al.}{2003}]{iwasawa03}
{Iwasawa} K.,  {Wilson} A.~S.,  {Fabian} A.~C.,   {Young} A.~J.,  2003, \mn@doi
  [\mnras] {10.1046/j.1365-8711.2003.06857.x}, \href
  {https://ui.adsabs.harvard.edu/abs/2003MNRAS.345..369I} {345, 369}

\bibitem[\protect\citeauthoryear{{Kakkad} et~al.,}{{Kakkad}
  et~al.}{2018}]{kakkad18}
{Kakkad} D.,  et~al., 2018, \mn@doi [\aap] {10.1051/0004-6361/201832790}, \href
  {https://ui.adsabs.harvard.edu/abs/2018A&A...618A...6K} {618, A6}

\bibitem[\protect\citeauthoryear{{Kubota} \& {Done}}{{Kubota} \&
  {Done}}{2018}]{Kubota_Done_18}
{Kubota} A.,  {Done} C.,  2018, \mn@doi [\mnras] {10.1093/mnras/sty1890}, \href
  {https://ui.adsabs.harvard.edu/abs/2018MNRAS.480.1247K} {480, 1247}

\bibitem[\protect\citeauthoryear{{Lauberts}}{{Lauberts}}{1982}]{lauberts1982}
{Lauberts} A.,  1982, {ESO/Uppsala survey of the ESO(B) atlas}

\bibitem[\protect\citeauthoryear{{Lepine} et~al.,}{{Lepine}
  et~al.}{2003}]{lepine+03}
{Lepine} J. R.~D.,  et~al., 2003, in {Iye} M.,  {Moorwood} A. F.~M.,  eds,
  Society of Photo-Optical Instrumentation Engineers (SPIE) Conference Series
  Vol. 4841, Instrument Design and Performance for Optical/Infrared
  Ground-based Telescopes. pp 1086--1095, \mn@doi{10.1117/12.461977}

\bibitem[\protect\citeauthoryear{{Lipovetsky}, {Neizvestny}  \&
  {Neizvestnaya}}{{Lipovetsky} et~al.}{1988}]{lipovetsky_1988}
{Lipovetsky} V.~A.,  {Neizvestny} S.~I.,   {Neizvestnaya} O.~M.,  1988,
  Soobshcheniya Spetsial'noj Astrofizicheskoj Observatorii, \href
  {https://ui.adsabs.harvard.edu/abs/1988SoSAO..55....5L} {55, 5}

\bibitem[\protect\citeauthoryear{{Lucy}}{{Lucy}}{1974}]{Lucy74}
{Lucy} L.~B.,  1974, \mn@doi [The Astronomical Journal] {10.1086/111605}, \href
  {https://ui.adsabs.harvard.edu/abs/1974AJ.....79..745L} {79, 745}

\bibitem[\protect\citeauthoryear{{Matt}, {Piro}, {Antonelli}, {Fink}, {Meurs}
  \& {Perola}}{{Matt} et~al.}{1994}]{matt94}
{Matt} G.,  {Piro} L.,  {Antonelli} L.~A.,  {Fink} H.~H.,  {Meurs} E.~J.~A.,
  {Perola} G.~C.,  1994, \aap, \href
  {https://ui.adsabs.harvard.edu/abs/1994A&A...292L..13M} {292, L13}

\bibitem[\protect\citeauthoryear{{May}, {Steiner}, {Ricci}, {Menezes}  \&
  {Andrade}}{{May} et~al.}{2016}]{DMay16}
{May} D.,  {Steiner} J.~E.,  {Ricci} T.~V.,  {Menezes} R.~B.,   {Andrade}
  I.~S.,  2016, \mn@doi [\mnras] {10.1093/mnras/stv2929}, \href
  {http://adsabs.harvard.edu/abs/2016MNRAS.457..949M} {457, 949}

\bibitem[\protect\citeauthoryear{{May}, {Rodr{\'{\i}}guez-Ardila}, {Prieto},
  {Fern{\'a}ndez-Ontiveros}, {Diaz}  \& {Mazzalay}}{{May}
  et~al.}{2018}]{DMay18}
{May} D.,  {Rodr{\'{\i}}guez-Ardila} A.,  {Prieto} M.~A.,
  {Fern{\'a}ndez-Ontiveros} J.~A.,  {Diaz} Y.,   {Mazzalay} X.,  2018, \mn@doi
  [\mnras] {10.1093/mnrasl/sly155}, \href
  {http://adsabs.harvard.edu/abs/2018MNRAS.481L.105M} {481, L105}

\bibitem[\protect\citeauthoryear{{Mazzalay} et~al.,}{{Mazzalay}
  et~al.}{2013}]{Mazzalay13}
{Mazzalay} X.,  et~al., 2013, \mn@doi [\mnras] {10.1093/mnras/sts204}, \href
  {http://adsabs.harvard.edu/abs/2013MNRAS.428.2389M} {428, 2389}

\bibitem[\protect\citeauthoryear{{Menezes}, {Steiner}  \& {Ricci}}{{Menezes}
  et~al.}{2014}]{Menezes14}
{Menezes} R.~B.,  {Steiner} J.~E.,   {Ricci} T.~V.,  2014, \mn@doi [\mnras]
  {10.1093/mnras/stt2381}, \href
  {http://adsabs.harvard.edu/abs/2014MNRAS.438.2597M} {438, 2597}

\bibitem[\protect\citeauthoryear{{Menezes}, {da Silva}, {Ricci}, {Steiner},
  {May}  \& {Borges}}{{Menezes} et~al.}{2015}]{Menezes15}
{Menezes} R.~B.,  {da Silva} P.,  {Ricci} T.~V.,  {Steiner} J.~E.,  {May} D.,
  {Borges} B.~W.,  2015, \mn@doi [\mnras] {10.1093/mnras/stv629}, \href
  {http://adsabs.harvard.edu/abs/2015MNRAS.450..369M} {450, 369}

\bibitem[\protect\citeauthoryear{{Menezes}, {Ricci}, {Steiner}, {da Silva},
  {Ferrari}  \& {Borges}}{{Menezes} et~al.}{2019}]{Menezes19}
{Menezes} R.~B.,  {Ricci} T.~V.,  {Steiner} J.~E.,  {da Silva} P.,  {Ferrari}
  F.,   {Borges} B.~W.,  2019, \mn@doi [\mnras] {10.1093/mnras/sty3337}, \href
  {https://ui.adsabs.harvard.edu/abs/2019MNRAS.483.3700M} {483, 3700}

\bibitem[\protect\citeauthoryear{{Morganti}, {Tsvetanov}, {Gallimore}  \&
  {Allen}}{{Morganti} et~al.}{1999}]{morganti99}
{Morganti} R.,  {Tsvetanov} Z.~I.,  {Gallimore} J.,   {Allen} M.~G.,  1999,
  \mn@doi [\aaps] {10.1051/aas:1999258}, \href
  {https://ui.adsabs.harvard.edu/abs/1999A&AS..137..457M} {137, 457}

\bibitem[\protect\citeauthoryear{{Morganti}, {Fogasy}, {Paragi}, {Oosterloo}
  \& {Orienti}}{{Morganti} et~al.}{2013a}]{morganti13b}
{Morganti} R.,  {Fogasy} J.,  {Paragi} Z.,  {Oosterloo} T.,   {Orienti} M.,
  2013a, \mn@doi [Science] {10.1126/science.1240436}, \href
  {https://ui.adsabs.harvard.edu/abs/2013Sci...341.1082M} {341, 1082}

\bibitem[\protect\citeauthoryear{{Morganti}, {Frieswijk}, {Oonk}, {Oosterloo}
  \& {Tadhunter}}{{Morganti} et~al.}{2013b}]{morganti13a}
{Morganti} R.,  {Frieswijk} W.,  {Oonk} R.~J.~B.,  {Oosterloo} T.,
  {Tadhunter} C.,  2013b, \mn@doi [\aap] {10.1051/0004-6361/201220734}, \href
  {https://ui.adsabs.harvard.edu/abs/2013A&A...552L...4M} {552, L4}

\bibitem[\protect\citeauthoryear{{Mukherjee}, {Bicknell}, {Sutherland}  \&
  {Wagner}}{{Mukherjee} et~al.}{2016}]{Muk16}
{Mukherjee} D.,  {Bicknell} G.~V.,  {Sutherland} R.,   {Wagner} A.,  2016,
  \mn@doi [\mnras] {10.1093/mnras/stw1368}, \href
  {http://adsabs.harvard.edu/abs/2016MNRAS.461..967M} {461, 967}

\bibitem[\protect\citeauthoryear{{Mukherjee}, {Wagner}, {Bicknell}, {Morganti},
  {Oosterloo}, {Nesvadba}  \& {Sutherland}}{{Mukherjee} et~al.}{2018}]{Murk18}
{Mukherjee} D.,  {Wagner} A.~Y.,  {Bicknell} G.~V.,  {Morganti} R.,
  {Oosterloo} T.,  {Nesvadba} N.,   {Sutherland} R.~S.,  2018, \mn@doi [\mnras]
  {10.1093/mnras/sty067}, \href
  {https://ui.adsabs.harvard.edu/abs/2018MNRAS.476...80M} {476, 80}

\bibitem[\protect\citeauthoryear{{M{\"u}ller-S{\'a}nchez}, {Prieto}, {Hicks},
  {Vives-Arias}, {Davies}, {Malkan}, {Tacconi}  \&
  {Genzel}}{{M{\"u}ller-S{\'a}nchez} et~al.}{2011}]{Muller11}
{M{\"u}ller-S{\'a}nchez} F.,  {Prieto} M.~A.,  {Hicks} E.~K.~S.,  {Vives-Arias}
  H.,  {Davies} R.~I.,  {Malkan} M.,  {Tacconi} L.~J.,   {Genzel} R.,  2011,
  \mn@doi [\apj] {10.1088/0004-637X/739/2/69}, \href
  {http://adsabs.harvard.edu/abs/2011ApJ...739...69M} {739, 69}

\bibitem[\protect\citeauthoryear{{Negrete}, {Dultzin}, {Marziani}  \&
  {Sulentic}}{{Negrete} et~al.}{2014}]{Negrete2014}
{Negrete} C.~A.,  {Dultzin} D.,  {Marziani} P.,   {Sulentic} J.~W.,  2014,
  \mn@doi [Advances in Space Research] {10.1016/j.asr.2013.11.037}, \href
  {https://ui.adsabs.harvard.edu/abs/2014AdSpR..54.1355N} {54, 1355}

\bibitem[\protect\citeauthoryear{{Osterbrock} \& {Ferland}}{{Osterbrock} \&
  {Ferland}}{2006}]{osterbrock06}
{Osterbrock} D.~E.,  {Ferland} G.~J.,  2006, {Astrophysics of gaseous nebulae
  and active galactic nuclei}

\bibitem[\protect\citeauthoryear{{Panda}}{{Panda}}{2021}]{Panda_2021}
{Panda} S.,  2021, \mn@doi [\aap] {10.1051/0004-6361/202140393}, \href
  {https://ui.adsabs.harvard.edu/abs/2021A&A...650A.154P} {650, A154}

\bibitem[\protect\citeauthoryear{{Panda}}{{Panda}}{2022}]{Panda2022}
{Panda} S.,  2022, \mn@doi [Frontiers in Astronomy and Space Sciences]
  {10.3389/fspas.2022.850409}, \href
  {https://ui.adsabs.harvard.edu/abs/2022FrASS...950409P} {9, 850409}

\bibitem[\protect\citeauthoryear{{Panda}, {Czerny}  \& {Wildy}}{{Panda}
  et~al.}{2017}]{Panda_etal_2017}
{Panda} S.,  {Czerny} B.,   {Wildy} C.,  2017, \mn@doi [Frontiers in Astronomy
  and Space Sciences] {10.3389/fspas.2017.00033}, \href
  {https://ui.adsabs.harvard.edu/abs/2017FrASS...4...33P} {4, 33}

\bibitem[\protect\citeauthoryear{{Panda}, {Czerny}, {Adhikari}, {Hryniewicz},
  {Wildy}, {Kuraszkiewicz}  \& {{\'S}niegowska}}{{Panda}
  et~al.}{2018}]{Panda_etal_2018}
{Panda} S.,  {Czerny} B.,  {Adhikari} T.~P.,  {Hryniewicz} K.,  {Wildy} C.,
  {Kuraszkiewicz} J.,   {{\'S}niegowska} M.,  2018, \mn@doi [\apj]
  {10.3847/1538-4357/aae209}, \href
  {https://ui.adsabs.harvard.edu/abs/2018ApJ...866..115P} {866, 115}

\bibitem[\protect\citeauthoryear{{Piconcelli}, {Bianchi}, {Vignali},
  {Jim{\'e}nez-Bail{\'o}n}  \& {Fiore}}{{Piconcelli}
  et~al.}{2011}]{Piconcelli11}
{Piconcelli} E.,  {Bianchi} S.,  {Vignali} C.,  {Jim{\'e}nez-Bail{\'o}n} E.,
  {Fiore} F.,  2011, \mn@doi [\aap] {10.1051/0004-6361/201117462}, \href
  {https://ui.adsabs.harvard.edu/abs/2011A&A...534A.126P} {534, A126}

\bibitem[\protect\citeauthoryear{{Prieto}, {Rodr{\'\i}guez-Ardila}, {Panda}  \&
  {Marinello}}{{Prieto} et~al.}{2022}]{Prieto_etal_2022}
{Prieto} A.,  {Rodr{\'\i}guez-Ardila} A.,  {Panda} S.,   {Marinello} M.,  2022,
  \mn@doi [\mnras] {10.1093/mnras/stab3414}, \href
  {https://ui.adsabs.harvard.edu/abs/2022MNRAS.510.1010P} {510, 1010}

\bibitem[\protect\citeauthoryear{{Proxauf}, {{\"O}ttl}  \&
  {Kimeswenger}}{{Proxauf} et~al.}{2014}]{Proxauf_Kimeswenger14}
{Proxauf} B.,  {{\"O}ttl} S.,   {Kimeswenger} S.,  2014, \mn@doi [\aap]
  {10.1051/0004-6361/201322581}, \href
  {https://ui.adsabs.harvard.edu/abs/2014A&A...561A..10P} {561, A10}

\bibitem[\protect\citeauthoryear{{Richardson}}{{Richardson}}{1972}]{Richardson72}
{Richardson} W.~H.,  1972, Journal of the Optical Society of America
  (1917-1983), \href {https://ui.adsabs.harvard.edu/abs/1972JOSA...62...55R}
  {62, 55}

\bibitem[\protect\citeauthoryear{{Riffel} et~al.,}{{Riffel}
  et~al.}{2022}]{riffel22}
{Riffel} R.,  et~al., 2022, \mn@doi [\mnras] {10.1093/mnras/stac740}, \href
  {https://ui.adsabs.harvard.edu/abs/2022MNRAS.512.3906R} {512, 3906}

\bibitem[\protect\citeauthoryear{{Rodr{\'\i}guez-Ardila} \&
  {Fonseca-Faria}}{{Rodr{\'\i}guez-Ardila} \&
  {Fonseca-Faria}}{2020}]{Rodriguez-Ardila_Fonseca-Faria20}
{Rodr{\'\i}guez-Ardila} A.,  {Fonseca-Faria} M.~A.,  2020, \mn@doi [\apjl]
  {10.3847/2041-8213/ab901b}, \href
  {https://ui.adsabs.harvard.edu/abs/2020ApJ...895L...9R} {895, L9}

\bibitem[\protect\citeauthoryear{{Rodr{\'{\i}}guez-Ardila}, {Prieto}, {Viegas}
  \& {Gruenwald}}{{Rodr{\'{\i}}guez-Ardila} et~al.}{2006}]{Ardila06}
{Rodr{\'{\i}}guez-Ardila} A.,  {Prieto} M.~A.,  {Viegas} S.,   {Gruenwald} R.,
  2006, \mn@doi [\apj] {10.1086/508864}, \href
  {http://adsabs.harvard.edu/abs/2006ApJ...653.1098R} {653, 1098}

\bibitem[\protect\citeauthoryear{{Rodr{\'{\i}}guez-Ardila}, {Prieto},
  {Mazzalay}, {Fern{\'a}ndez-Ontiveros}, {Luque}  \&
  {M{\"u}ller-S{\'a}nchez}}{{Rodr{\'{\i}}guez-Ardila} et~al.}{2017}]{Ardila17}
{Rodr{\'{\i}}guez-Ardila} A.,  {Prieto} M.~A.,  {Mazzalay} X.,
  {Fern{\'a}ndez-Ontiveros} J.~A.,  {Luque} R.,   {M{\"u}ller-S{\'a}nchez} F.,
  2017, \mn@doi [\mnras] {10.1093/mnras/stx1401}, \href
  {http://adsabs.harvard.edu/abs/2017MNRAS.470.2845R} {470, 2845}

\bibitem[\protect\citeauthoryear{{Rose}, {Elvis}  \& {Tadhunter}}{{Rose}
  et~al.}{2015}]{Rose15}
{Rose} M.,  {Elvis} M.,   {Tadhunter} C.~N.,  2015, \mn@doi [\mnras]
  {10.1093/mnras/stv113}, \href
  {https://ui.adsabs.harvard.edu/abs/2015MNRAS.448.2900R} {448, 2900}

\bibitem[\protect\citeauthoryear{{Schmitt} \& {Storchi-Bergmann}}{{Schmitt} \&
  {Storchi-Bergmann}}{1995}]{Schmitt_Storchi-Bergmann95}
{Schmitt} H.~R.,  {Storchi-Bergmann} T.,  1995, \mn@doi [\mnras]
  {10.1093/mnras/276.2.592}, \href
  {https://ui.adsabs.harvard.edu/abs/1995MNRAS.276..592S} {276, 592}

\bibitem[\protect\citeauthoryear{{Smith} \& {Wilson}}{{Smith} \&
  {Wilson}}{2001}]{smith01}
{Smith} D.~A.,  {Wilson} A.~S.,  2001, \mn@doi [\apj] {10.1086/321667}, \href
  {https://ui.adsabs.harvard.edu/abs/2001ApJ...557..180S} {557, 180}

\bibitem[\protect\citeauthoryear{{V{\'e}ron-Cetty} \&
  {V{\'e}ron}}{{V{\'e}ron-Cetty} \& {V{\'e}ron}}{2006}]{Veron06}
{V{\'e}ron-Cetty} M.~P.,  {V{\'e}ron} P.,  2006, \mn@doi [\aap]
  {10.1051/0004-6361:20065177}, \href
  {https://ui.adsabs.harvard.edu/abs/2006A&A...455..773V} {455, 773}

\bibitem[\protect\citeauthoryear{{Wandel}, {Peterson}  \& {Malkan}}{{Wandel}
  et~al.}{1999}]{Wandel1999}
{Wandel} A.,  {Peterson} B.~M.,   {Malkan} M.~A.,  1999, \mn@doi [\apj]
  {10.1086/308017}, \href
  {https://ui.adsabs.harvard.edu/abs/1999ApJ...526..579W} {526, 579}

\bibitem[\protect\citeauthoryear{{Zovaro}, {Sharp}, {Nesvadba}, {Bicknell},
  {Mukherjee}, {Wagner}, {Groves}  \& {Krishna}}{{Zovaro}
  et~al.}{2019}]{zovaro19}
{Zovaro} H. R.~M.,  {Sharp} R.,  {Nesvadba} N. P.~H.,  {Bicknell} G.~V.,
  {Mukherjee} D.,  {Wagner} A.~Y.,  {Groves} B.,   {Krishna} S.,  2019, \mn@doi
  [\mnras] {10.1093/mnras/stz233}, \href
  {https://ui.adsabs.harvard.edu/abs/2019MNRAS.484.3393Z} {484, 3393}

\makeatother
\end{thebibliography}




\appendix

\section{Supplementary diagnostic plots from {\sc cloudy} modelling}

\begin{figure}
    \centering
    \includegraphics[width=\columnwidth]{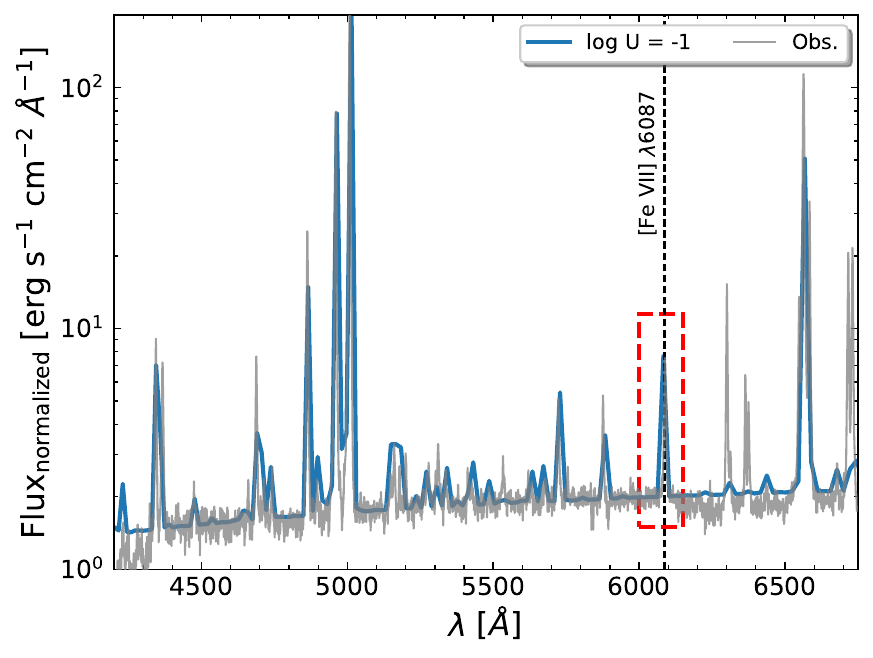}
    \caption{Similar to the left panel in Figure \ref{fig:compare-spectrum} (for the nuclear region) but for ionisation parameter, log $U$ = -1. This particular modelled spectrum (in blue) reproduces the observed [Fe\,{\sc vii}]$\lambda$6087 \AA~ emission (in grey) with the lowest residual.}
    \label{fig:cloudy_fe7}
\end{figure}

\begin{figure*}
    \centering
    \includegraphics[width=\textwidth]{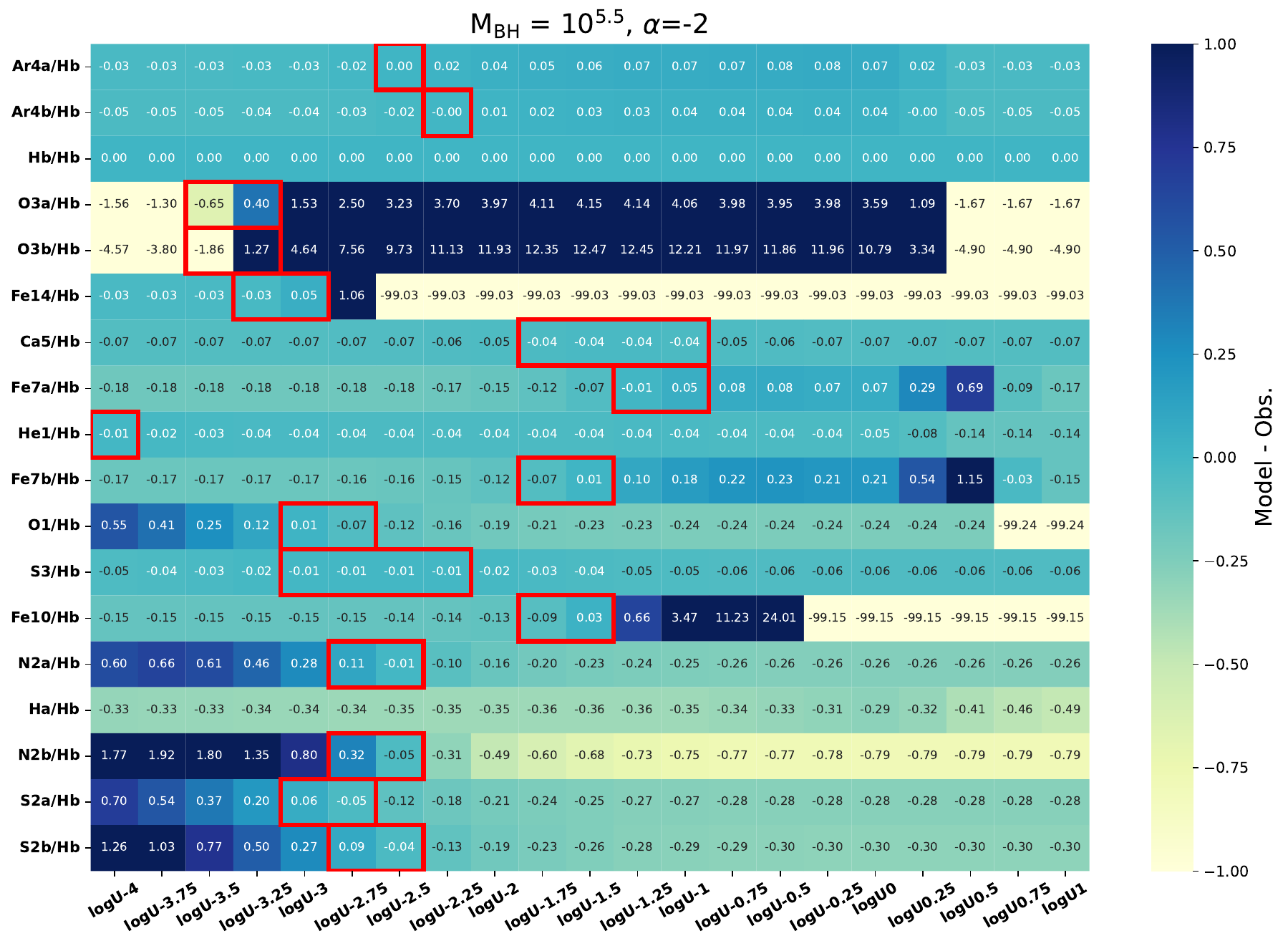}
    \caption{Similar to Figure \ref{fig:heatmap-1}, but for the alternate cases M$_{\rm BH}$ = 10$^{5.50}$ M$_{\odot}$ for the density law slope, $\alpha$ = -2.}
    \label{fig:heatmap-alt1}
\end{figure*}
    
\begin{figure*}
\includegraphics[width=\textwidth]{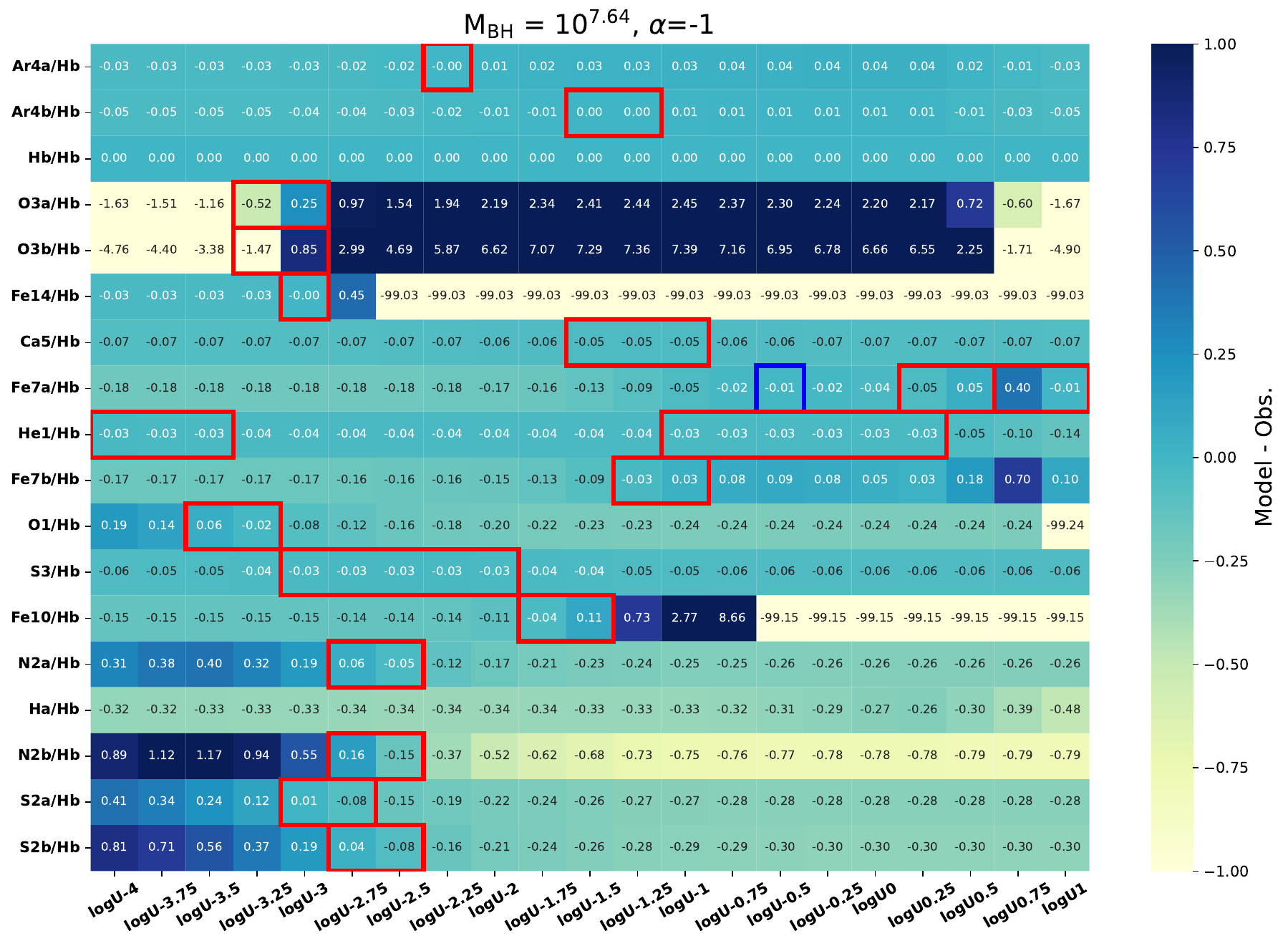}
    \caption{Similar to Figure \ref{fig:heatmap-1}, but for the alternate cases M$_{\rm BH}$ = 10$^{7.64}$ M$_{\odot}$ for the density law slope, $\alpha$ = -1. The value marked in blue is the second-to-best solution for the Fe {\sc vii}/H$\beta$ intensity ratio.}
    \label{fig:heatmap-alt2}
\end{figure*}

\begin{figure*}
\includegraphics[width=\textwidth]{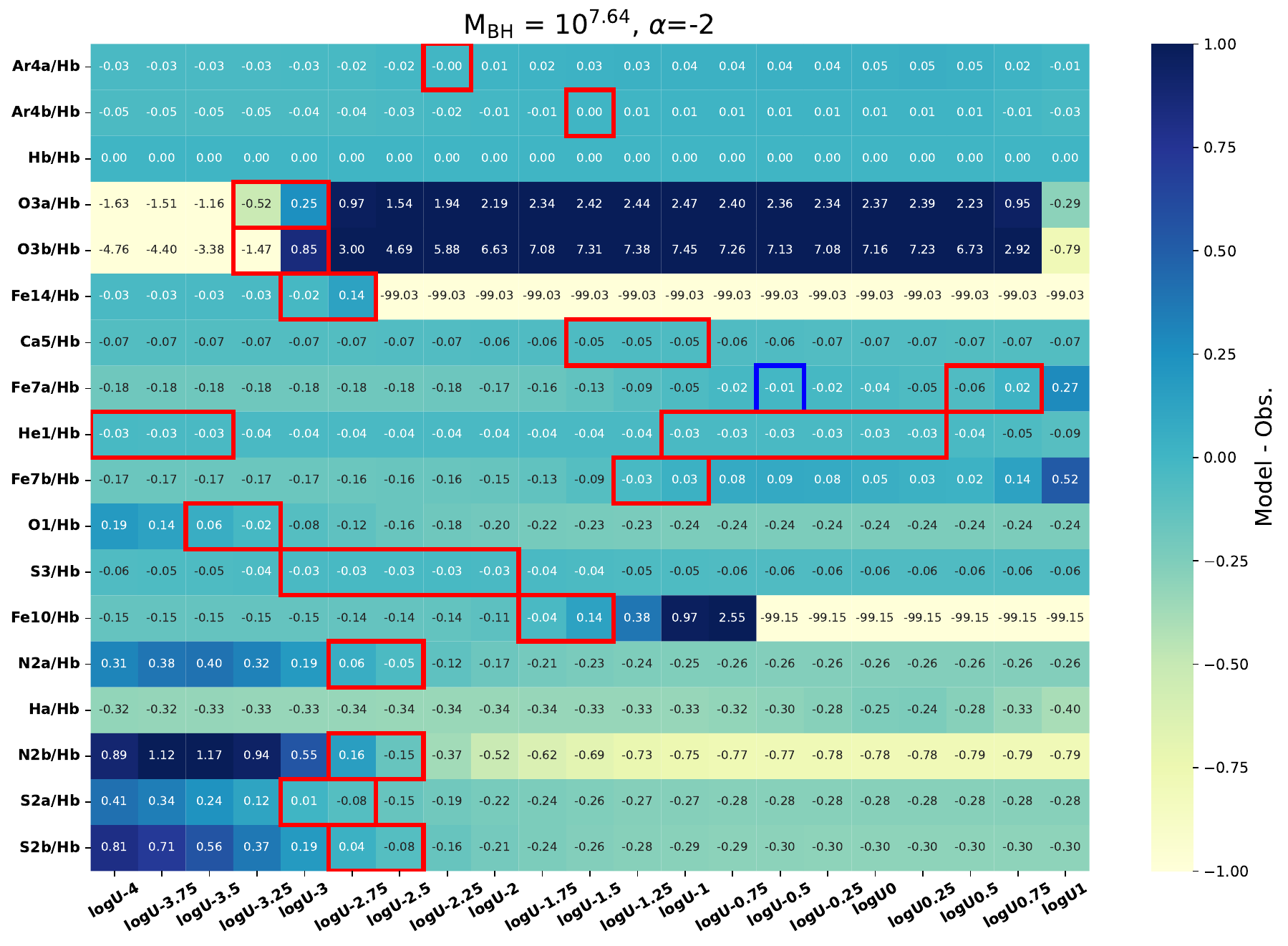}
    
    \caption{Similar to Figure \ref{fig:heatmap-1}, but for the alternate cases M$_{\rm BH}$ = 10$^{7.64}$ M$_{\odot}$ for the density law slope, $\alpha$ = -2. The value marked in blue is the second-to-best solution for the Fe {\sc vii}/H$\beta$ intensity ratio.}
    \label{fig:heatmap-alt3}    
\end{figure*}


\bsp	
\label{lastpage}
\end{document}